\documentclass[10pt,journal,twocolumn]{IEEEtran} 
\usepackage{wrapfig,booktabs,fancyhdr,amsmath,amsfonts,tabularx,numprint}
\usepackage{cite,bm,bbm,amssymb,amsthm,url,multirow,times,enumitem,comment}
\usepackage{mathtools,siunitx,balance,tikz,adjustbox,graphicx,array}
\usepackage{algorithm}
\usepackage{algorithmic}
\usepackage{amsthm}
\usepackage[font=footnotesize]{subcaption}
\usepackage[colorlinks=true, linkcolor=black, citecolor=blue, urlcolor=blue]{hyperref}
\usepackage[switch,pagewise]{lineno}
\usepackage{acro}
\usepackage{eqparbox}

\newcommand{\sfrac}[2]{\textstyle \frac{#1}{#2}}
\newcommand{\ba}{\begin{array}}
\newcommand{\ea}{\end{array}}

\renewcommand{\Re}{\mathbb{R}}

\DeclareMathAlphabet{\mathpzc}{OT1}{pzc}{m}{it}
\DeclareMathOperator*{\argmin}{\arg\!\min}
\DeclareMathOperator*{\argmax}{\arg\!\max}

\acsetup{single}

\newcommand{\fc}{\ensuremath{f_\text{c}}}
\newcommand{\Ta}{\ensuremath{T_{\text{a}}}}
\newcommand{\Na}{\ensuremath{N_{\text{a}}}}
\newcommand{\Ts}{\ensuremath{T_{\text{s}}}}

\newcommand{\Kpm}{\ensuremath{\mathcal{K}_{m}^{\text{P}}}}
\newcommand{\Kdm}{\ensuremath{\mathcal{K}_{m}^{\text{D}}}}
\newcommand{\K}{\ensuremath{\mathcal{K}}}
\newcommand{\Ktm}{\ensuremath{\tilde{\mathcal{K}}_{m}}}

\newcommand{\cs}{\ensuremath{c_{\text{sym}}}}

\DeclareSIUnit \belm {Bm}

\DeclareAcronym{awgn}{short = AWGN, long = additive white Gaussian noise}
\DeclareAcronym{ici}{short = ICI, long = intercarrier interference}
\DeclareAcronym{isi}{short = ISI, long = intersymbol interference}
\DeclareAcronym{sinr}{short = SINR, long = signal-to-interference-and-noise ratio}
\DeclareAcronym{snr}{short = SNR, long = signal-to-noise ratio}
\DeclareAcronym{ofdm}{short = OFDM, long = orthogonal frequency-division multiplexing}
\DeclareAcronym{mimo}{short = MIMO, long = multiple-input multiple-output}
\DeclareAcronym{simo}{short = SIMO, long = single-input multiple-output}
\DeclareAcronym{siso}{short = SISO, long = single-input single-output}
\DeclareAcronym{cfo}{short = CFO, long =  carrier frequency offset}
\DeclareAcronym{cpe}{short = CPE, long = common phase error}
\DeclareAcronym{lte}{short = LTE, long = long term evolution}
\DeclareAcronym{nr}{short = NR, long = new radio}
\DeclareAcronym{crlb}{short = CRB, long = Cramer-Rao bound}
\DeclareAcronym{mcrlb}{short = MCRB, long = Modified Cramer-Rao bound}
\DeclareAcronym{acf}{short = ACF, long = autocorrelation function}
\DeclareAcronym{zzb}{short = ZZB, long = Ziv-Zakai bound}
\DeclareAcronym{mmse}{short = MMSE, long = minimum mean square error}
\DeclareAcronym{lmmse}{short = LMMSE, long = linear minimum mean square error}
\DeclareAcronym{rmse}{short = RMSE, long = root mean square error}
\DeclareAcronym{mrc}{short = MRC, long = maximum-ratio combining}
\DeclareAcronym{toa}{short = TOA, long = time-of-arrival}
\DeclareAcronym{pdf}{short = PDF, long = probability density function}
\DeclareAcronym{cdf}{short = CDF, long = cumulative distribution function}
\DeclareAcronym{ccdf}{short = CCDF, long = complementary cumulative distribution function}
\DeclareAcronym{dft}{short = DFT, long = discrete Fourier transform}
\DeclareAcronym{jcas}{short = JCAS, long = joint communications and sensing}
\DeclareAcronym{isac}{short = ISAC, long = integrated sensing and communications}
\DeclareAcronym{prs}{short = PRS, long = positioning reference signal}
\DeclareAcronym{sdr}{short = SDR, long = software-defined radio}
\DeclareAcronym{gnss}{short = GNSS, long = global navigation satellite system}
\DeclareAcronym{tdoa}{short = TDOA, long = time-difference-of-arrival}
\DeclareAcronym{rtt}{short = RTT, long = round-trip-time}
\DeclareAcronym{aoa}{short = AOA, long = angle-of-arrival}
\DeclareAcronym{aod}{short = AOD, long = angle-of-departure}
\DeclareAcronym{leo}{short = LEO, long = low-Earth-orbit}
\DeclareAcronym{ml}{short = ML, long = maximum likelihood}
\DeclareAcronym{enu}{short = ENU, long = east-north-up}
\DeclareAcronym{ntn}{short = NTN, long = non-terrestrial-network}
\DeclareAcronym{dd}{short = DD, long = decision-directed}
\DeclareAcronym{nda}{short = NDA, long = non-data-aided}
\DeclareAcronym{los}{short = LOS, long = line-of-sight}
\DeclareAcronym{sop}{short = SOP, long = signal-of-opportunity, long-plural-form = signals-of-opportunity}
\DeclareAcronym{sage}{short = SAGE, long = space-alternating generalized expectation-maximization}
\DeclareAcronym{rnti}{short = RNTI, long = Radio Network Temporary Identifier}
\DeclareAcronym{ekf}{short = EKF, long = extended Kalman filter}

\begin{document}
\title{OFDM-Based Positioning with Unknown Data Payloads: Bounds and Applications to LEO PNT}

\author{Andrew M. Graff, \IEEEmembership{Graduate Student Member, IEEE} and Todd E. Humphreys, \IEEEmembership{Senior Member, IEEE}
	\thanks{A. Graff is with the Department of Electrical and Computer Engineering, The University of Texas at Austin, Austin, TX 78712, USA \mbox{(e-mail: andrewgraff@utexas.edu)}.}
	\thanks{T. Humphreys is with the Department of Aerospace Engineering and Engineering Mechanics, The University of Texas at Austin, Austin, TX 78712, USA \mbox{(e-mail: todd.humphreys@utexas.edu)}.}
}

\maketitle
\begin{abstract}
	This paper presents bounds, estimators, and signal design strategies for exploiting both known pilot resources and unknown data payload resources in \ac{toa}-based positioning systems with \ac{ofdm} signals. It is the first to derive the \ac{zzb} on \ac{toa} estimation for \ac{ofdm} signals containing both known pilot and unknown data resources. In comparison to the \acp{crlb} derived in prior work, this \ac{zzb} captures the low-\ac{snr} thresholding effects in \ac{toa} estimation and accounts for an unknown carrier phase. The derived \ac{zzb} is evaluated against \acp{crlb} and empirical \ac{toa} error variances. It is then evaluated on signals with resource allocations optimized for pilot-only \ac{toa} estimation, quantifying the performance gain over the best-case pilot-only signal designs. Finally, the positioning accuracy of maximum-likelihood and decision-directed estimators is evaluated on simulated \acl{leo} \acl{ntn} channels and compared against their respective \acp{zzb}.
\end{abstract}


\newif\ifpreprint
\preprinttrue

\ifpreprint

\pagestyle{plain}
\thispagestyle{fancy}  
\fancyhf{} 
\renewcommand{\headrulewidth}{0pt}
\rfoot{\footnotesize \bf This work has been submitted to the IEEE for possible publication. Copyright may\\be transferred without notice, after which this version may no longer be accessible.}
\lfoot{\footnotesize \bf
	Copyright \copyright~2024 by Andrew M. Graff \\ and Todd E. Humphreys}

\else

\thispagestyle{empty}
\pagestyle{empty}

\fi

\section{Introduction}
\label{sec:unknown_data:intro}
\acresetall

\IEEEPARstart{P}{ositioning} services within existing wireless communications networks are becoming an increasingly important source of accurate localization. Such services can provide exceptional accuracy due to their access to large bandwidths and widespread deployment, making them an attractive alternative to traditional \ac{gnss} positioning. \Ac{leo} \acp{ntn} in particular are a prime candidate for accurate user positioning services because of their ability to provide near-global coverage and strong signals with exceptionally large bandwidths \cite{dureppagari2023ntn,iannucci2022fusedLeo}. These \ac{leo} \acp{ntn} are undergoing a massive expansion in satellite deployment \cite{lin2021path}, further increasing their viability as the new standard for global positioning while simultaneously providing high-throughput communications.

An overwhelming majority of communications networks, including the 5G \ac{ntn} standard \cite{lin2021path}, operate through \ac{ofdm}, which divides the spectrum into time and frequency resource elements that may be allocated with either known pilot resources or unknown data resources. Since the structure of these pilot resources is known to users in the network, users may obtain \ac{toa} estimates by correlating their received signal against the known pilot resources. These \ac{toa} estimates may then be used in positioning protocols such as pseudorange multilateration, \ac{tdoa}, or \ac{rtt} \cite{dwivedi2021positioning}. But this scheme creates a challenging design tradeoff:  allocating additional pilot resources improves positioning accuracy at the expense of data throughput, since \ac{ofdm} resources must be diverted from data \cite{graff2024purposeful}. This tradeoff becomes especially complex for satellites in \acp{ntn}, which must additionally manage their power consumption when balancing the tradeoff between communications and positioning services \cite{iannucci2022fusedLeo}. Such a ``zero-sum game'' poses problems for rapidly expanding \acp{ntn} tasked with handling an ever-growing demand for both increased positioning accuracy and higher data rates. However, a new and enticing scheme emerges if users exploit both pilot resources and data resources in their \ac{toa} estimation.

Two approaches exist for users to exploit data resources in their \ac{toa} estimation: \ac{dd} and \ac{ml} estimation. The decision-directed estimator makes hard decoding decisions on the unknown data and then correlates the received signal against both the pilot resources and decoded data to improve estimation accuracy. While \ac{dd} estimators are efficient and effective at high \acp{snr}, they are prone to decoding errors. Although these decoding errors may be mitigated through error correcting codes, such codes may not be usable for data resources intended for other users since networks may intentionally obfuscate the coding from other users; witness the scrambling by the \ac{rnti} in \ac{lte} and 5G \ac{nr}. In contrast to \ac{dd} estimators, \ac{ml} estimators, commonly referred to as \ac{nda} estimators in prior work when no known pilots are used \cite{bellili2010cramer,masmoudi2011non}, do not make hard decoding decisions but instead evaluate the likelihood of data resources over all symbols in the constellation. Since an overwhelming majority of the spectrum in communications networks is allocated to data resources, these \ac{dd} and \ac{ml} estimators can harness a much greater amount of signal power compared to the pilot-only estimator, resulting in significantly reduced \ac{toa} estimation errors.

When evaluating signals for \ac{toa} estimation, it is crucial to consider the impact that resource allocation has on the \ac{toa} likelihood function, which exhibits a mainlobe centered around the true \ac{toa}, sidelobes located away from the true \ac{toa} outside the mainlobe, and grating lobes caused by aliasing. At high \ac{snr}, \ac{toa} estimation errors will be concentrated near the true \ac{toa} within the mainlobe. At low \ac{snr}, however, \ac{toa} estimates may latch on to sidelobes, significantly increasing estimation error variance \cite{zeira1994realizable,nanzer2016bandpass,sahinoglu2008ultra}. Bounds such as the Barankin bound \cite{barankin1949,mcaulay1971barankin} and \ac{zzb} \cite{Ziv1969} were derived to capture the behavior of this thresholding effect, which is ignored by the simpler \ac{crlb}. When only pilot resources are used in \ac{toa} estimation, the likelihood function is closely related to the \ac{acf}. However, the inclusion of unknown data resources in the likelihood function alters its shape in a complex manner, potentially introducing new sidelobes and sharpening the mainlobe. A bound that captures both the effects of unknown data and low \ac{snr} is needed to accurately characterize the positioning performance of \ac{dd} and \ac{ml} estimators.

\ac{leo} \acp{ntn} are especially amenable to positioning with unknown data resources. The large satellite constellation sizes increase the likelihood of a \ac{los} path between users and satellites. Furthermore, phased arrays at both ends provide exceptional multipath mitigation. Finally, \ac{leo} satellites may be able to pre-compensate for Doppler due to their highly directive beams and small cell size. As a result, the fading in the post-beamforming channels remains exceptionally flat across wide bandwidths --- as large as \SI{240}{\mega\hertz} in the case of Starlink \cite{humphreys2023starlinkSignalStructure}. Under these favorable channels conditions, users only need to estimate and compensate for the \ac{toa} and carrier phase to equalize the received signal and begin decoding data.

This paper derives the \ac{zzb} on \ac{toa} estimation error variance for \ac{ofdm} signals containing both pilot resources and unknown data resources. The derived \ac{zzb} is then compared against \acp{crlb} derived in prior work and against empirical \ac{toa} estimation error variance. Empirical \ac{toa} estimates are obtained with both \ac{ml} and \ac{dd} estimators. Three variants of \ac{ml} estimators are considered: one that exploits only pilot resources, another based on only data resources, and a third that harnesses both pilot and data resources. These will be referred to as the pilot-only, data-only, and pilot-plus-data \ac{ml} estimators, respectively. Candidate resource allocations are then generated over a range of \acp{snr} that optimize the placement of \acp{prs} in the frequency domain to minimize the pilot-only \ac{toa} \ac{zzb}. The pilot-plus-data \ac{toa} \ac{zzb} is then evaluated on these optimized signals to quantify the reduction in \ac{toa} estimation error that can be achieved over the optimal allocations for pilot-only estimation. Finally, \ac{leo} \ac{ntn} channels are simulated for a satellite constellation servicing a single cell. The positioning accuracy of a user in the serviced cell is evaluated using Monte Carlo methods for the pilot-only \ac{ml}, data-only \ac{ml}, pilot-plus-data \ac{ml}, and \ac{dd} estimators. These results are then compared against the derived \ac{zzb}.

\subsection{Prior Work}

Prior work has studied several \ac{toa}-based positioning algorithms with \ac{ofdm} signals. Algorithms such as \ac{tdoa}, pseudorange multilateration, and \ac{rtt} are supported within the existing 5G \ac{nr} standards \cite{dwivedi2021positioning}. More advanced approaches have demonstrated accurate positioning with 5G \ac{nr} signals using both \ac{toa} and \ac{aoa} measurements in an \ac{ekf} \cite{koivisto2017joint} and using multipath parameter estimates obtained from signals with optimized beam power allocations \cite{kakkavas2021power}. Alternatively, \ac{lte} \cite{shamaei2018lte} and 5G \ac{nr} \cite{shamaei2021receiver,neinavaie2021cognitive} signals have been used as \acp{sop} for positioning, a paradigm that requires no cooperation between the user and the network. While exceptional positioning accuracy is demonstrated in this prior work, both the network-supported and \ac{sop} methods only obtain position estimates from known reference signals embedded in the \ac{ofdm} signal and do not exploit the vast quantity of unknown data resources that are present in typical \ac{lte} and 5G \ac{nr} downlink signals. The presence of \ac{ofdm} reference signals has also been detected in a cognitive manner for \ac{sop} positioning \cite{neinavaie2021cognitive,neinavaie2023cognitive}, but this approach still relies on the allocation of reference signals by the network.

The analysis of positioning within communications networks has been extended to the context of \ac{leo} \acp{ntn}. Scheduling for \ac{leo} constellations providing both communications and positioning services has been analyzed in \cite{iannucci2022fusedLeo}. The authors in \cite{xv2023joint} optimized \ac{leo} beamforming and beam scheduling to minimize the user positioning \ac{crlb}. This work demonstrates the potential improvements \ac{leo} positioning services may provide over existing \ac{gnss} solutions. However, prior work has not yet analyzed \ac{ofdm} resource allocation for \ac{leo} positioning nor explored the potential for exploiting data resources in \ac{toa} estimation.


Outside of positioning, \ac{dd} and \ac{nda} estimation have been thoroughly studied in the context of communications.
Prior work has analyzed \ac{dd} estimators for \ac{ofdm} frequency-offset estimation \cite{Shi2005}, \ac{ofdm} channel estimation in high-velocity channels \cite{ran2003decision}, and \ac{mimo} channel tracking \cite{karami2006decision}.
The authors in \cite{xiong2024data} propose a \ac{dd} channel estimator for overcoming pilot contamination in cell-free massive \ac{mimo} networks.
This body of work demonstrates the potential improvements in estimator accuracy that may be gained through hard decoding decisions on unknown data.
Similarly, prior work has analyzed \ac{nda} estimators for \ac{ofdm} frequency-offset estimation \cite{ma2001non}, \ac{ofdm} timing recovery \cite{al2006novel}, and \ac{ofdm} \ac{snr} estimation \cite{socheleau2008non}.
Prior work has also studied the \acp{crlb} of \ac{nda} estimation.
Bellili et al. derived a \ac{crlb} for \ac{nda} time \cite{masmoudi2011closed} and frequency \cite{bellili2010cramer} estimation for square-QAM constellations that is tighter than the simpler \ac{mcrlb}.
The authors in \cite{masmoudi2017nda} also propose a \ac{nda} \ac{toa} estimator that uses importance sampling to reduce computational costs, and they compare its performance against the \ac{mcrlb}.

Whereas prior work has extensively studied \ac{nda} and \ac{dd} estimators for communications purposes, only a limited body of work has studied their applicability to positioning.
The authors in \cite{monfared2020iterative} proposed a \ac{nda} \ac{aoa} estimator for positioning with Gaussian frequency-shift-keying signals.
Wang et al. proposed a semiblind \ac{ofdm} range tracker which, after initialization with known pilot resources, tracks the multipath components of the channel using \ac{dd} decoding decisions and a Kalman filter  \cite{wang2015semiblind}.
Similarly, the authors in \cite{adam2013semi} proposed a semiblind channel estimator for positioning that improves its estimates of the multipath components of the channel using \ac{dd} decoding, comparing performance against the \ac{mcrlb}.
Mensing et al. proposed a \ac{dd} \ac{toa} estimator for \ac{tdoa} positioning with intercell interference \cite{mensing2009dd}.
This work demonstrates how unknown data payloads can be exploited, either through \ac{nda} or \ac{dd} estimation, to improve positioning accuracy within communications networks. However, these papers do not compare \ac{nda} and \ac{dd} estimators against one another to understand how each estimator's errors change with \ac{snr}. Furthermore, \cite{monfared2020iterative} and \cite{adam2013semi} did not consider \ac{ofdm} signals, and \cite{monfared2020iterative} did not consider \ac{toa}-based positioning. Finally, \cite{monfared2020iterative,adam2013semi,mensing2009dd} compared estimator accuracy only against the \ac{crlb}, which is unable to capture low-\ac{snr} thresholding effects.

To overcome the limitations of the \ac{crlb}, several studies have used the \ac{zzb} for analyzing positioning performance in \ac{ofdm} systems with pilot-only \ac{toa} estimation. Prior work has used the \ac{zzb} to characterize the \ac{toa} precision of different parameterizations of \ac{ofdm} pilot resource allocations \cite{graff2024purposeful,laas2021ziv,dammann2016optimizing,staudinger2017optimized}. Furthermore, the \ac{zzb} has been used as an optimization criteria to solve for \ac{ofdm} pilot resource allocations that minimize \ac{toa} estimation errors \cite{graff2024ziv}. The \ac{zzb} on direct position estimation has also been derived in \cite{gusi2018ziv}. However, prior work has not derived the \ac{zzb} in the context of unknown data or for \ac{nda} estimation.

\subsection{Contributions}
The main contributions of this paper are as follows:
\begin{itemize}
	\item A novel derivation of the \ac{zzb} on \ac{toa} estimation error variance for \ac{ofdm} signals with unknown data resources. This novel bound is compared against the \ac{crlb}, \ac{mcrlb}, and empirical errors from Monte-Carlo simulation.
	\item A comparison of the pilot-only and pilot-plus-data \acp{zzb} for \ac{ofdm} signals with resources optimized for pilot-only \ac{toa} estimation. This comparison provides insights into the potential gains achieved by exploiting unknown data and informs how pilot resources can be allocated to minimize overhead while meeting \ac{toa} accuracy requirements.
	\item Evaluation of the empirical positioning errors achieved by both \ac{ml} and \ac{dd} estimators on simulated \ac{leo} satellite channels in comparison to the \ac{zzb}.
\end{itemize}

The remainder of this paper is organized as follows. Section~\ref{sec:unknown_data:signal_model} introduces the signal model. Section~\ref{sec:unknown_data:toa_bounds} defines the \ac{toa} \acp{crlb} and derives the \ac{toa} \ac{zzb} for \ac{ofdm} signals with payloads containing unknown data. Section~\ref{sec:unknown_data:ml_estimation} defines the \ac{ml} estimators for the pilot-only, data-only, and pilot-plus-data cases as well as the \ac{dd} estimator. Section~\ref{sec:unknown_data:results_bounds} compares the derived \ac{zzb} against the \acp{crlb} and Monte Carlo \ac{toa} errors. Section~\ref{sec:unknown_data:results_prs} evaluates the pilot-plus-data \ac{zzb} on signals with resource allocations optimized for pilot-only estimation. Section~\ref{sec:unknown_data:results_leo} evaluates the positioning accuracy of the pilot-only \ac{ml}, data-only \ac{ml}, pilot-plus-data \ac{ml}, and \ac{dd} estimators on simulated \ac{leo} \ac{ntn} channels, comparing them against the derived \ac{zzb}. Finally, Section~\ref{sec:unknown_data:conclusion} closes the paper by drawing conclusions from the results.

\textbf{Notation:} Column vectors are denoted with lowercase bold, e.g., $\bm{x}$. Matrices are denoted with uppercase bold, e.g., $\bm{X}$. Scalars are denoted without bold, e.g., $x$. The $i$th entry of a vector $\bm{x}$ is denoted $x[i]$ or in shorthand as $x_i$. The Euclidean norm is denoted $||\bm{x}||$. The cardinality of a set $\mathcal{S}$ is denoted $|\mathcal{S}|$. Real transpose is represented by the superscript $T$ and conjugate transpose by the superscript $H$. The Q-function is denoted as $Q(\cdot)$. Zero-based indexing is used throughout the paper; e.g., $x[0]$ refers to the first element of $\bm{x}$.

\section{Signal Model}
\label{sec:unknown_data:signal_model}

Consider an \ac{ofdm} signal with $K$ subcarriers, $N_{\text{sym}}$ symbols, a subcarrier spacing of $\Delta_{\text{f}}\;\SI{}{\hertz}$, and a payload $x_{m}[k]$ for symbol indices $m \in \mathcal{M}$ and subcarrier indices $k \in \K$, where $\mathcal{M} = \{0,1,\ldots, N_{\text{sym}}-1\}$ and $\K = \{0,1,\ldots, K-1\}$. Let $d[k]$ be the mapping from subcarrier indices to offsets in frequency from the carrier in units of subcarriers. This map is defined as $d[k] = k$ for $k = 0,1,\ldots,\sfrac{K}{2}-1$ and $d[k] = k-K$ for $k = \sfrac{K}{2}, \sfrac{K}{2}+1,\ldots, K-1$. This signal propagates through a doubly-selective channel at a carrier frequency $\fc$ with baseband frequency-domain channel coefficients $h_{m}[k]$ and experiences a \ac{los} time delay $\tau$, phase shift $\phi$, and \ac{awgn} $v_{m}[k] \sim \mathcal{CN}(0,\sigma^2)$. Assuming negligible \acl{ici} due to small Doppler and negligible \acl{isi} due to a sufficiently long cyclic-prefix, the baseband received signal $y_{m}[k]$ is modeled in the frequency domain as
\begin{align}
	y_{m}[k] &= \alpha_{m}[k] x_{m}[k] + v_{m}[k],\\
	\alpha_{m}[k] &= h_{m}[k]\exp\left(-j2\pi d[k] \Delta_{\text{f}} \tau + j\phi\right).
	\label{eq:unknown_data:sig_model}
\end{align}
If the channel coefficients are constant across frequency and time, the complex gain $\alpha_{m}[k]$ can be instead modeled as
\begin{align}
	\alpha_{m}[k] &= \sqrt{g}\exp\left(-j2\pi d[k] \Delta_{\text{f}} \tau + j\phi\right),
	\label{eq:unknown_data:sig_model_flat}
\end{align}
where $g$ is the channel gain. The bounds and estimators of this paper are derived under the frequency-flat time-invariant model in (\ref{eq:unknown_data:sig_model_flat}), while the simulated channels in the results use the frequency-selective time-varying model in (\ref{eq:unknown_data:sig_model}).

The payload $x_{m}[k]$ may contain either pilot resources, data resources, or be empty. Define $\Kpm$ as the set of subcarrier indices containing pilot resources, $\Kdm$ as the set of subcarrier indices containing data resources, and $\Ktm \triangleq \Kpm \bigcup \Kdm$ as their union, during symbol $m$. Each data resource is modeled as randomly selected from a symbol constellation $\mathcal{C}$ with uniform probability and statistical independence from all other resource elements. This is expressed as $x_{m}[k] = c_{m}[k]$ and $P(c_{m}[k] = \cs) = \frac{1}{|\mathcal{C}|}$ for $m \in \mathcal{M}$, $k \in \Kdm$, and $\cs \in \mathcal{C}$. Furthermore, the constellation is modeled as having unit average power such that $\mathbb{E}\left[|x_{m}[k]|^2\right] = 1$.

\section{TOA Estimation Error Bounds}
\label{sec:unknown_data:toa_bounds}

Let $\hat{\tau}$ be an unbiased estimate of the true \ac{toa} $\tau$ and define $\gamma_{m}[k] = \sfrac{|\alpha_{m}[k]|^2}{\sigma^2}$ as the \ac{snr} at subcarrier $k$ during symbol $m$. This section will define bounds on the \ac{toa} estimation error variance $\mathbb{E}\left[(\hat{\tau} - \tau)^2\right] $ under the simplified signal model in (\ref{eq:unknown_data:sig_model_flat}).

\subsection{Cramer-Rao Bounds}

The simplest bound is the \ac{crlb} when $\hat{\tau}$ is estimated using only pilot resources, which takes the form \cite{xu2016maximum}
\begin{align}
	\mathbb{E}\left[(\hat{\tau} - \tau)^2\right] &\geq \sigma_{\text{CRLB,P}}^2 = I_{\text{CRLB,P}}^{-1}\\
	I_{\text{CRLB,P}}&\triangleq 8\pi^2 \Delta_{\text{f}}^2 \sum_{m \in \mathcal{M}} \sum_{k \in \Kpm} d^{2}[k] \gamma_{m}[k].
	\label{eq:unknown_data:pilot_crlb}
\end{align}

In comparison to the pilot-only \ac{crlb}, the derivation of the \ac{crlb} for the data-only estimator is more difficult since the likelihood function for each resource element is a Gaussian mixture distribution function with each Gaussian centered at each symbol in the constellation $\mathcal{C}$. One simplifying approach is to derive the \ac{mcrlb} \cite{d1994modified} by conditioning on the unknown symbols. This simplifies to a form similar to the pilot-only \ac{crlb}
\begin{align}
	\mathbb{E}\left[(\hat{\tau} - \tau)^2\right] &\geq \sigma_{\text{MCRLB}}^2 = I_{\text{MCRLB}}^{-1}\\
I_{\text{MCRLB}}&\triangleq 8\pi^2 \Delta_{\text{f}}^2 \sum_{m \in \mathcal{M}} \sum_{k \in \Kdm} d^{2}[k] \gamma_{m}[k].
\label{eq:unknown_data:mcrlb}
\end{align}

A tighter \ac{crlb} for unknown data is derived in \cite{bellili2010cramer} without using the \ac{mcrlb}. While this bound was derived for frequency estimation, it is easily mapped to \ac{toa} estimation with \ac{ofdm} signals. This \ac{crlb} takes the form
\begin{align}
	&\mathbb{E}\left[(\hat{\tau} - \tau)^2\right] \geq \sigma_{\text{CRLB,D}}^2 = I_{\text{CRLB,D}}^{-1},
\end{align}
where
\begin{align}
	I_{\text{CRLB,D}} \triangleq 8 \pi^2  \Delta_{\text{f}}^2 \sum_{\substack{m \in \mathcal{M}\\k \in \Kdm}} &\left(\frac{1+\gamma_{m}[k]}{A_2}\psi(\gamma_{m}[k]) - \gamma_{m}[k]\right)\nonumber\\
	&\times d^{2}[k] \gamma_{m}[k],
	\label{eq:unknown_data:bellili_crlb}
\end{align}
and $\psi(\cdot)$ and $A_2$ are defined in \cite[Eqs. (37)-(38)]{bellili2010cramer}.

These \acp{crlb} are useful for analyzing the high-\ac{snr} precision of \ac{toa} estimators, and the \ac{crlb} from \cite{bellili2010cramer} importantly captures the increased errors caused by uncertainty in the symbol selected from the constellation, making the bound in (\ref{eq:unknown_data:bellili_crlb}) tighter than the \ac{mcrlb} in (\ref{eq:unknown_data:mcrlb}). These fundamental error bounds can provide valuable insights into the efficiency of estimators and can serve as an optimization criteria for evaluating different signal designs and resource allocations. However, the \acp{crlb} ignore the impact that sidelobes in the likelihood function have on estimation error, making these bounds inapt at lower \acp{snr} \cite{graff2024purposeful}.

\subsection{Ziv-Zakai Bounds}

%

The \ac{zzb} is superior to the \ac{crlb} for \ac{toa} estimation analysis because it captures the low-\ac{snr} thresholding effects caused by sidelobes. The bound considers a binary detection problem with equally-likely hypotheses: (1) the received signal experienced delay $\tau_0$ and phase $\phi_0$, and (2) the received signal experienced delay $\tau_0 + \tau_{1}$ and phase $\phi_0 + \phi_{1} \pmod{2\pi}$. Define $\bm{\theta} \triangleq [\tau,\; \phi]^T$, $\bm{\theta}_0 \triangleq [\tau_0,\; \phi_0]^T$, and $\bm{\theta}_{1} \triangleq [\tau_{0}{+}\tau_{1},\; \phi_{0}{+}\phi_{1} \pmod{2\pi}]^T$. $\bm{\theta}$ is treated as a random variable with a known \textit{a priori} distribution. The hypothesis test can be expressed as
\begin{align}
	\frac{p(\bm{y}|\bm{\theta}{=}\bm{\theta}_{0})}{p(\bm{y}|\bm{\theta}{=}\bm{\theta}_{1})} \underset{\mathcal{H}_{1}}{\overset{\mathcal{H}_{0}}{\gtrless}} \nu, \quad \quad \genfrac{}{}{0pt}{}{\mathcal{H}_{0} \; : \; \bm{\theta} = \bm{\theta}_{0}}{\mathcal{H}_{1} \; : \; \bm{\theta} = \bm{\theta}_{1}},
	\label{eq:unknown_data:hypothesis}
\end{align}
where $\bm{y}$ is the vector containing all $y_{m}[k]$ for $m \in \mathcal{M}$ and $k \in \Ktm$, and $\nu$ is the detection threshold. The \ac{zzb} is concerned with the minimum error probability of this hypothesis test, which corresponds to a threshold of $\nu=1$ when $\mathcal{H}_{0}$ and $\mathcal{H}_{1}$ are assumed equally-likely.

As in \cite{graff2024ziv}, the time delays will be normalized by the \ac{ofdm} sampling period $\Ts = \sfrac{1}{K \Delta_{\text{f}}}$, creating $z \triangleq \sfrac{\tau}{\Ts}$, $z_0  \triangleq \sfrac{\tau_0}{\Ts}$ and $z_{1} \triangleq \sfrac{\tau_{1}}{\Ts}$. The minimum error probability of this detection problem is assumed to be shift-invariant, allowing the hypotheses to be simplified without loss of generality by assuming $\tau_0 = 0$ and $\phi_0 = 0$. Then the log of the likelihood ratio in (\ref{eq:unknown_data:hypothesis}) can be denoted
\begin{align}
	\log \Lambda(\bm{y},\bm{\theta}_{1}) \triangleq \log{p(\bm{y}|\bm{\theta}{=}\bm{\theta}_0) } - \log{p(\bm{y}|\bm{\theta}{=}\bm{\theta}_{1}) },
	\label{eq:unknown_data:llr_1}
\end{align}
and the minimum error probability of the hypothesis test is defined as the probability that the log-likelihood ratio in (\ref{eq:unknown_data:llr_1}), conditioned on $\bm{\theta}=\bm{\theta}_{0}$, is less than zero:
\begin{align}
	P_{\text{min}}(z_{1},\phi_{1}) \triangleq P\left(\log \Lambda(\bm{y},\bm{\theta}_{1}) < 0\right|\bm{\theta}{=}\bm{\theta}_0).
	\label{eq:unknown_data:pmin_1}
\end{align}

Define $\bm{R}_{\bm{\theta}}$ as the estimation error covariance of $\bm{\theta}$ and $\mathcal{V}\{\cdot\}$ as the valley-filling function \cite{bellini1974bounds}. Assuming \textit{a priori} knowledge that the \ac{toa} is uniformly distributed on $[0,\Ta]$ and the phase is uniformly distributed on $[0,2\pi]$, and noting the scale-invariance of the valley-filling function, the \ac{zzb} on \ac{toa} error variance can be defined as \cite{bell1997extended}
\begin{align}
	&\mathbb{E}\left[(\hat{\tau} - \tau)^2\right] = \bm{a}^T \bm{R}_{\bm{\theta}} \bm{a}  \label{eq:unknown_data:zzb_vec}\\
	&\geq \sigma_{\text{ZZB}}^2 \nonumber\\
	&\triangleq \frac{1}{\Ta}\int_{0}^{\Ta} \tau_{1} \mathcal{V}\{(\Ta - \tau_{1}) \max_{\phi_{1}} \left[P_{\text{min}}(\tau_{1}/T_s, \phi_{1})\right]\} d\tau_{1} \nonumber\\
	&= \frac{\Ts^2}{\Na}\int_{0}^{\Na} z_{1} \mathcal{V}\{(\Na - z_{1}) \max_{\phi_{1}} \left[P_{\text{min}}(z_{1},\phi_{1})\right]\} dz_{1},\nonumber
\end{align}
where $\Na \triangleq \sfrac{\Ta}{\Ts}$.

Expressions for $P_{\text{min}}(z_{1},\phi_{1})$ will now be derived. The likelihood of $\bm{y}$ is
\begin{align}
	p(\bm{y}|\bm{\theta}) = \prod_{m \in \mathcal{M}} \prod_{k \in \Ktm} p(y_{m}[k]|\bm{\theta}),
\end{align}
which follows from the independence of each resource element. The log-likelihood ratio in (\ref{eq:unknown_data:llr_1}) can be expressed as
\begin{align}
	&\log \Lambda(\bm{y},\bm{\theta}_{1}) \nonumber\\
	&= \sum_{m \in \mathcal{M}} \sum_{k \in \Ktm} \log{p(y_{m}[k]|\bm{\theta}{=}\bm{\theta}_0)} - \log{p(y_{m}[k]|\bm{\theta}{=}\bm{\theta}_{1})} \nonumber\\
	&= \sum_{m \in \mathcal{M}} \sum_{k \in \Ktm} \log \Lambda(y_{m}[k],\bm{\theta}_{1}),
	\label{eq:unknown_data:llr}
\end{align}
where $\log \Lambda(y_{m}[k],\bm{\theta}_{1}) \triangleq \log{p(y_{m}[k]|\bm{\theta}{=}\bm{\theta}_0)} - \log{p(y_{m}[k]|\bm{\theta}{=}\bm{\theta}_{1})}$ is the log-likelihood ratio for the resource at symbol $m$ and subcarrier $k$. Similarly, the minimum error probability in (\ref{eq:unknown_data:pmin_1}) can be expressed as
\begin{align}
	&P_{\text{min}}(z_{1},\phi_{1}) \label{eq:unknown_data:p_err}\\
	&= P\left(\sum_{m \in \mathcal{M}} \sum_{k \in \Ktm} \log \Lambda(y_{m}[k],\bm{\theta}_{1}) < 0 \Bigg| \bm{\theta}{=}\bm{\theta}_0\right).\nonumber
\end{align}
To derive this probability, the distribution of $\log \Lambda$ conditioned on the parameter vector $\bm{\theta}=\bm{\theta}_0$ must be analyzed. Accordingly, all expectations through the remainder of this section are conditioned on $\bm{\theta}=\bm{\theta}_{0}$.

\subsubsection*{Unknown Data Resources}
First consider the case of unknown data resources. The likelihood of $y_{m}[k]$ conditioned on the parameter vector $\bm{\theta}$ and conditioned on knowledge of the symbol $c_{m}[k]$ is
\begin{align}
	&p(y_{m}[k]|c_{m}[k],\bm{\theta}) \\
	&= \frac{1}{\pi \sigma^2} \exp{\left(\frac{-1}{\sigma^2}\left|y_{m}[k]-\mu_{m}[k]\nu_{k}(\bm{\theta})\right|^2 \right)}, \nonumber
\end{align}
where $\nu_{k}(\bm{\theta}) \triangleq \exp{\left(-j2\pi z d[k] / K + j\phi\right)}$ and $\mu_{m}[k] \triangleq \sqrt{g} c_{m}[k]$. Assuming equally-likely symbols in the constellation, the likelihood of $y_{m}[k]$ is given by
\begin{align}
	p(y_{m}[k]|\bm{\theta}) = \frac{1}{|\mathcal{C}|}\sum_{\cs \in \mathcal{C}} p(y_{m}[k]|c_{m}[k] = \cs,\bm{\theta}),
\end{align}
which is a Gaussian mixture distribution function. The log-likelihood of $y_{m}[k]$ then becomes
\begin{align}
	&\log p(y_{m}[k]|\bm{\theta}) \label{eq:unknown_data:log_likelihood}\\
	&= \log{\frac{1}{|\mathcal{C}|}} + \log{\sum_{\cs \in \mathcal{C}} p(y_{m}[k]|c_{m}[k] = \cs, \bm{\theta})} \nonumber\\
	&= \log{\frac{1}{\pi\sigma^2|\mathcal{C}|}} + \log{\sum_{\cs \in \mathcal{C}} \exp{\left(\frac{-1}{\sigma^2}\left|y_{m}[k]-\mu_{m}[k]\nu_{k}(\bm{\theta}) \right|^2\right)}} \nonumber\\
	&= \log{\frac{1}{\pi\sigma^2|\mathcal{C}|}} - \frac{1}{\sigma^2} |y_{m}[k]|^2 \nonumber\\
	&+ \log\sum_{\cs \in \mathcal{C}} \exp\left(\frac{1}{\sigma^2}\left(2 \Re\{y^{\ast}_{m}[k]\mu_{m}[k]\nu_{k}(\bm{\theta})\} - g|\cs|^2\right)\right). \nonumber
\end{align}
With this log-likelihood defined, $\log p(y_{m}[k]|\bm{\theta}{=}\bm{\theta}_0)$ and $\log p(y_{m}[k]|\bm{\theta}{=}\bm{\theta}_{1})$ can be substituted into (\ref{eq:unknown_data:llr}), resulting in
\begin{align}
	&\log \Lambda(y_{m}[k],\bm{\theta}_{1}) \label{eq:unknown_data:llr_unknown}\\
	&= \log\sum_{\cs \in \mathcal{C}} \exp\left(\frac{1}{\sigma^2}\left(2 \Re\{y^{\ast}_{m}[k]\mu_{m}[k]\} - g|\cs|^2\right)\right) \nonumber\\
	&- \log\sum_{\cs \in \mathcal{C}} \exp\left(\frac{1}{\sigma^2}\left(2 \Re\{y^{\ast}_{m}[k]\mu_{m}[k]\nu_{k}(\bm{\theta}_{1})\} - g|\cs|^2\right)\right). \nonumber
\end{align}

The distribution of this log-sum-exp form in (\ref{eq:unknown_data:llr_unknown}) is difficult to analyze. Conditioned on a specific symbol $c_{m}[k]$, $y_{m}[k]$ becomes Gaussian distributed and (\ref{eq:unknown_data:llr_unknown}) becomes the difference of the log of two lognormal sums. Prior work has approximated the log of lognormal sums as Gaussian-distributed \cite{mehta2007approximating}. Likewise, the log-likelihood ratio in (\ref{eq:unknown_data:llr_unknown}) will be approximated as a Gaussian distribution matching the first and second moments.

Since the log-likelihood is a function of a Gaussian random variable conditioned on knowledge of the symbol $c_{m}[k]$, its moment generating function is easily expressed as
\begin{align}
	&M_{\log \Lambda}(t) \triangleq \mathbb{E}\left[\exp{\left(t \log \Lambda(y_{m}[k],\bm{\theta}_{1})\right)}|\bm{\theta}{=}\bm{\theta}_{0}\right]\\
	&= \mathbb{E}\left[\mathbb{E}\left[\exp{\left(t \log \Lambda(y_{m}[k],\bm{\theta}_{1})\right)}|c_{m}[k],\bm{\theta}{=}\bm{\theta}_{0}\right]\right] \nonumber\\
	&= \frac{1}{|\mathcal{C}|} \sum_{\cs \in \mathcal{C}} \mathbb{E}\left[\exp{\left(t \log \Lambda(y_{m}[k],\bm{\theta}_{1})\right)}|c_{m}[k]=\cs,\bm{\theta}{=}\bm{\theta}_{0}\right],\nonumber
\end{align}
where the smoothing property allows the expectation to be conditioned on the symbol $c_{m}[k]$. The inner expectation is taken over the noise $v_{m}[k]$. The first moment can then be computed as
\begin{align}
	&\mathbb{E}\left[\log \Lambda(y_{m}[k],\bm{\theta}_{1})|\bm{\theta}{=}\bm{\theta}_{0}\right] = \frac{\partial}{\partial t} M_{\log \Lambda}(t) \bigg\rvert_{t=0} \label{eq:unknown_data:mom_1}\\
	&= \frac{1}{|\mathcal{C}|} \sum_{\cs \in \mathcal{C}} \mathbb{E}\left[\log \Lambda(y_{m}[k],\bm{\theta}_{1}) | c_{m}[k]=\cs,\bm{\theta}{=}\bm{\theta}_{0}\right], \nonumber
\end{align}
and the second moment can be computed as
\begin{align}
	&\mathbb{E}\left[(\log \Lambda(y_{m}[k],\bm{\theta}_{1}))^2|\bm{\theta}{=}\bm{\theta}_{0}\right] = \frac{\partial^2}{\partial t^2} M_{\log \Lambda}(t) \bigg\rvert_{t=0} \label{eq:unknown_data:mom_2}\\
	&= \frac{1}{|\mathcal{C}|} \sum_{\cs \in \mathcal{C}} \mathbb{E}\left[\left(\log \Lambda(y_{m}[k],\bm{\theta}_{1})\right)^2 | c_{m}[k]=\cs,\bm{\theta}{=}\bm{\theta}_{0}\right]. \nonumber
\end{align}


Since these expectations are taken over a complex Gaussian distribution, they can be approximated using a Gauss-Hermite quadrature \cite[25.4.46]{abramowitz1968handbook}. Consider a Gauss-Hermite quadrature of size $N$ with weights $w[n]$ and nodes $\delta_{v}[n]$ for $n \in 0,1,\ldots, N-1$. Then define $\tilde{\delta}_{v}[n_{\text{I}},n_{\text{Q}}] \triangleq \delta_{v}[n_{\text{I}}]+j\delta_{v}[n_{\text{Q}}]$. Note that conditioned on $c_{m}[k] = \cs$, the expected value of $y_{m}[k]$ is $\mu_{m}[k] = \sqrt{g} \cs$. The expectation in (\ref{eq:unknown_data:mom_1}) can then be expressed as
\begin{align}
	&\mathbb{E}\left[\log \Lambda(y_{m}[k],\bm{\theta}_{1}) | c_{m}[k]=\cs,\bm{\theta}{=}\bm{\theta}_{0}\right] \\
	&\approx \sum_{n_{\text{I}}=0}^{N-1} \sum_{n_{\text{Q}}=0}^{N-1} w[n_{\text{I}}] w[n_{\text{Q}}] \log \Lambda \left(\mu_{m}[k]+\tilde{\delta}_{v}[n_{\text{I}},n_{\text{Q}}],\bm{\theta}_{1}\right). \nonumber
\end{align}
Similarly, the expectation in (\ref{eq:unknown_data:mom_2}) can be expressed as
\begin{align}
	&\mathbb{E}\left[(\log \Lambda(y_{m}[k],\bm{\theta}_{1}))^2 | c_{m}[k]=\cs,\bm{\theta}{=}\bm{\theta}_{0}\right] \\
	&\approx \sum_{n_{\text{I}}=0}^{N-1} \sum_{n_{\text{Q}}=0}^{N-1} w[n_{\text{I}}] w[n_{\text{Q}}] \left(\log \Lambda \left(\mu_{m}[k]+\tilde{\delta}_{v}[n_{\text{I}},n_{\text{Q}}],\bm{\theta}_{1}\right)\right)^2. \nonumber
\end{align}
Finally, the variance of the log-likelihood ratio for the resource at symbol $m$ and subcarrier $k$ is
\begin{align}
	&\text{Var}\left(\log \Lambda(y_{m}[k],\bm{\theta}_{1})|\bm{\theta}{=}\bm{\theta}_{0}\right) =\\
	&\mathbb{E}\left[(\log \Lambda(y_{m}[k],\bm{\theta}_{1}))^2|\bm{\theta}{=}\bm{\theta}_{0}\right] - \mathbb{E}\left[\log \Lambda(y_{m}[k],\bm{\theta}_{1})|\bm{\theta}{=}\bm{\theta}_{0}\right]^2. \nonumber
\end{align}
The mean and variance of the log-likelihood ratio for unknown data resources is now quantified.

\subsubsection*{Known Pilot Resources}

Now consider the case of known pilot resources. The log-likelihood ratio takes the form
\begin{align}
	&\log \Lambda(y_{m}[k],\bm{\theta}_{1}) \label{eq:unknown_data:llr_pilot}\\
	&= \log{p(y_{m}[k]|\bm{\theta}{=}\bm{\theta}_0)} - \log{p(y_{m}[k]|\bm{\theta}{=}\bm{\theta}_{1})} \nonumber\\
	&= \frac{1}{\sigma^2}\left(\left|y_{m}[k]-\mu_{m}[k]\nu_{k}(\bm{\theta}_{1})\right|^2 - \left|y_{m}[k]-\mu_{m}[k]\right|^2\right) \nonumber\\
	&= \frac{2}{\sigma^2}\left( \Re\{y_{m}^{\ast}[k]\mu_{m}[k]\} - \Re\{y_{m}^{\ast}[k]\mu_{m}[k]\nu_{k}(\bm{\theta}_{1})\}\right) \nonumber\\
	&= \frac{2}{\sigma^2}\Re\{y_{m}^{\ast}[k]\mu_{m}[k]\left(1 - \nu_{k}(\bm{\theta}_{1})\right)\}. \nonumber
\end{align}

Since (\ref{eq:unknown_data:llr_pilot}) is a linear function of a Gaussian random variable $y_{m}[k]$, the log-likelihood ratio is itself Gaussian distributed. The mean of the log-likelihood ratio can be expressed as
%
\begin{align}
	&\mathbb{E}\left[\log \Lambda(y_{m}[k],\bm{\theta}_{1})|\bm{\theta}{=}\bm{\theta}_{0}\right] \\
	&= \frac{2}{\sigma^2} |\mu_{m}[k]|^2 \left(1 - \cos\left(2\pi z_{1} d[k]/K + \phi_{1}\right)\right) \nonumber\\
	&= \frac{2\alpha^{2}_{m}[k]}{\sigma^2} |c_{m}[k]|^2 \left(1 - \cos\left(2\pi z_{1} d[k]/K + \phi_{1}\right)\right), \nonumber
\end{align}
and the variance of the log-likelihood ratio can be expressed as
%
\begin{align}
	&\text{Var}\left(\log \Lambda(y_{m}[k],\bm{\theta}_{1})|\bm{\theta}{=}\bm{\theta}_{0}\right) \\
	&= \frac{4}{\sigma^2}|\mu_{m}[k]|^2 \left(1 - \cos\left(2\pi z_{1} d[k]/K + \phi_{1}\right)\right). \nonumber
\end{align}
The probability of error using only pilot resources simplifies to the form seen in \cite{dardari2009ziv}. However, expressing the mean and variance of the log-likelihood ratio allows pilot resources and data resources to be combined together in the \ac{zzb}.


\subsubsection*{Unified Expression for $P_{\text{min}}(z,\phi)$}

Now that expressions have been derived for the mean and variance of the log-likelihood ratio at each resource element for both unknown data resources and known pilot resources, the distribution of the sum log-likelihood ratio $\log \Lambda(\bm{y},\bm{\theta}_{1})$ can be quantified. By approximating the log-likelihood ratio of the data resources as Gaussian, it follows that the sum log-likelihood is also approximately Gaussian. Furthermore, for large numbers of subcarriers, this approximation will improve by the central limit theorem.
The mean of the sum log-likelihood ratio $\log \Lambda(\bm{y},\bm{\theta}_{1})$ is
\begin{align}
	&\mathbb{E}\left[\log \Lambda(\bm{y},\bm{\theta}_{1})|\bm{\theta}{=}\bm{\theta}_{0}\right] \\
	&= \sum_{m \in \mathcal{M}} \sum_{k \in \Ktm} \mathbb{E}\left[\log \Lambda(y_{m}[k],\bm{\theta}_{1})|\bm{\theta}{=}\bm{\theta}_{0}\right], \nonumber
\end{align}
and its variance is
\begin{align}
	&\text{Var}\left(\log \Lambda(\bm{y},\bm{\theta}_{1})|\bm{\theta}{=}\bm{\theta}_{0}\right) \\
	&= \sum_{m \in \mathcal{M}} \sum_{k \in \Ktm} \text{Var}\left(\log \Lambda(y_{m}[k],\bm{\theta}_{1})|\bm{\theta}{=}\bm{\theta}_{0}\right). \nonumber
\end{align}
Finally, the probability of error in (\ref{eq:unknown_data:p_err}) can be expressed as
\begin{align}
	&P_{\text{min}}(z_{1},\phi_{1}) \label{eq:unknown_data:p_err_final}\\
	&\approx Q\left(\mathbb{E}\left[\log \Lambda(\bm{y},\bm{\theta}_{1})|\bm{\theta}{=}\bm{\theta}_{0}\right]/\sqrt{\text{Var}\left(\log \Lambda(\bm{y},\bm{\theta}_{1})|\bm{\theta}{=}\bm{\theta}_{0}\right)}\right), \nonumber
\end{align}
which can be substituted into (\ref{eq:unknown_data:zzb_vec}) to compute the \ac{zzb}. When only pilot resources are used, the probability of error in (\ref{eq:unknown_data:p_err_final}) simplifies to the form in \cite{dardari2009ziv}.


\section{Maximum Likelihood Estimation}
\label{sec:unknown_data:ml_estimation}

In the results presented in the following section, the bounds in Section~\ref{sec:unknown_data:toa_bounds} will be evaluated against \ac{ml} estimators and a \ac{dd} estimator. Recall that three variants of \ac{ml} estimator are considered: the pilot-only, data-only, and pilot-plus-data \ac{ml} estimators. The pilot-only \ac{ml} estimator can be expressed as
\begin{align}
	\hat{\bm{\theta}}_{\text{P}} = \argmax_{\bm{\theta}_{1}} \sum_{m \in \mathcal{M}} \sum_{k \in \Kpm} \log p(y_{m}[k]|\bm{\theta}{=}\bm{\theta}_{1}).
	\label{eq:unknown_data:ml_p}
\end{align}
The data-only \ac{ml} estimator is expressed as
\begin{align}
	\hat{\bm{\theta}}_{\text{D}} = \argmax_{\bm{\theta}_{1}} \sum_{m \in \mathcal{M}} \sum_{k \in \Kdm} \log p(y_{m}[k]|\bm{\theta}{=}\bm{\theta}_{1}).
	\label{eq:unknown_data:ml_d}
\end{align}
And the pilot-plus-data \ac{ml} estimator is expressed as
\begin{align}
	\hat{\bm{\theta}} = \argmax_{\bm{\theta}_{1}} \sum_{m \in \mathcal{M}} \sum_{k \in \Ktm} \log p(y_{m}[k]|\bm{\theta}{=}\bm{\theta}_{1}).
	\label{eq:unknown_data:ml_both}
\end{align}
Note that (\ref{eq:unknown_data:ml_p})-(\ref{eq:unknown_data:ml_both}) differ in the sets of subcarriers in the summation.

The \ac{dd} estimator requires an initial estimate of the signal parameters $\bm{\theta}$ and therefore will only be applied when both pilot and data resources are present in the signal. After obtaining the pilot-only \ac{ml} parameter estimate $\hat{\bm{\theta}}_{\text{P}}$, \ac{ml} decoding decisions are made:
\begin{align}
	\hat{c}_{m}[k] = \argmax_{\cs \in \mathcal{C}} \log p(y_{m}[k]|c_{m}[k]=\cs,\bm{\theta}{=}\hat{\bm{\theta}}_{\text{P}}).
	\label{eq:unknown_data:dd_decode}
\end{align}
It is important to note that this \ac{dd} estimator uses no error correction codes, which, as mentioned in Section~\ref{sec:unknown_data:intro}, cannot be assumed to be of benefit in the current context because they are user-specific whereas this paper's scheme is designed to exploit all data. After decoding the data resources, another \ac{ml} parameter estimate is obtained using both pilot and data resources by treating the decoded data resources as known symbols:
\begin{align}
	\hat{\bm{\theta}}_{\text{DD}} &= \argmax_{\bm{\theta}_{1}} \sum_{m \in \mathcal{M}} \sum_{k \in \Kpm} \log p(y_{m}[k]|\bm{\theta}{=}\bm{\theta}_{1}) \label{eq:unknown_data:ml_dd}\\
	 &+ \sum_{m \in \mathcal{M}} \sum_{k \in \Kdm} \log p(y_{m}[k]|c_{m}[k] =\hat{c}_{m}[k],\bm{\theta}{=}\bm{\theta}_{1}).\nonumber
\end{align}

The $\argmax$ operations in (\ref{eq:unknown_data:ml_p})-(\ref{eq:unknown_data:ml_both}) and (\ref{eq:unknown_data:ml_dd}) are evaluated using a grid search with quadratic peak interpolation. The \ac{toa} $z$ has a discretized grid spanning from $0$ to $N_a$ in intervals of size $\Delta_{z}$, while the phase $\phi$ has a discretized grid spanning from $0$ to $2\pi$ in intervals of size $\Delta_{\phi}$.

\section{Results}
\label{sec:unknown_data:results}

\subsection{Bounds}
\label{sec:unknown_data:results_bounds}

\begin{figure}[h]
	\centering
	\includegraphics[width=0.9\linewidth]{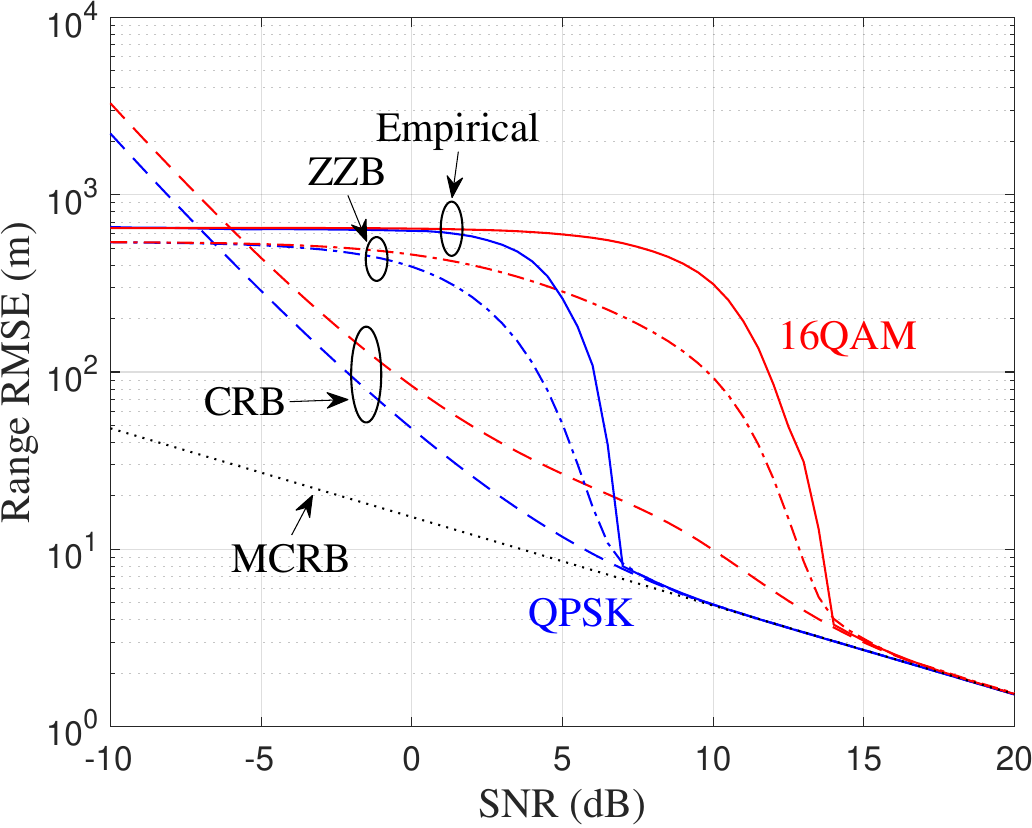}
	\caption{A comparison of the Monte Carlo empirical \acp{rmse}, \ac{mcrlb}, \ac{crlb}, and \ac{zzb} on \ac{toa} estimation using \textit{a priori} unknown data. Results are shown for both QPSK and 16QAM constellations.}
	\label{fig:unknown_data:bound_comparison}
\end{figure}

\begin{figure}[h]
	\centering
	\includegraphics[width=0.9\linewidth]{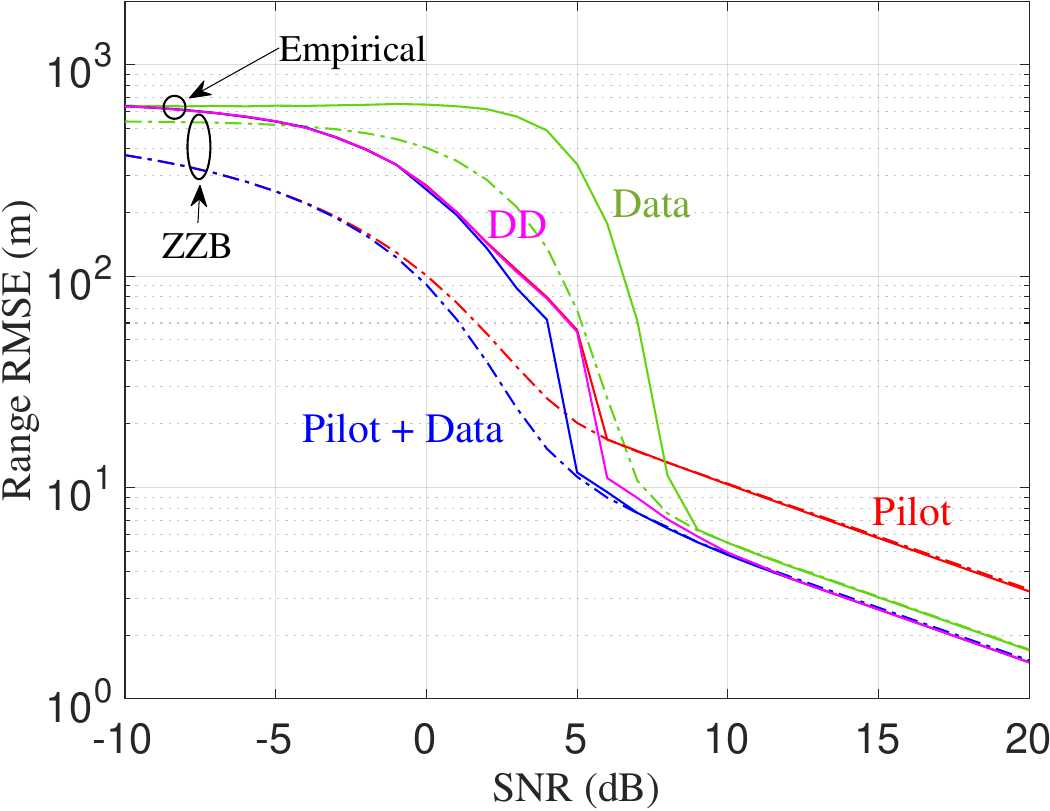}
	\caption{A comparison of the Monte Carlo empirical \acp{rmse} and bounds for pilot-only, data-only, and pilot-plus-data estimation. Results are shown for a single \ac{ofdm} symbol with $64$ subcarriers, QPSK modulation, and $8$ sparsely-placed pilot resources.}
	\label{fig:unknown_data:bound_comparison2}
\end{figure}

\begin{figure}[h]
	\centering
	\includegraphics[width=0.9\linewidth]{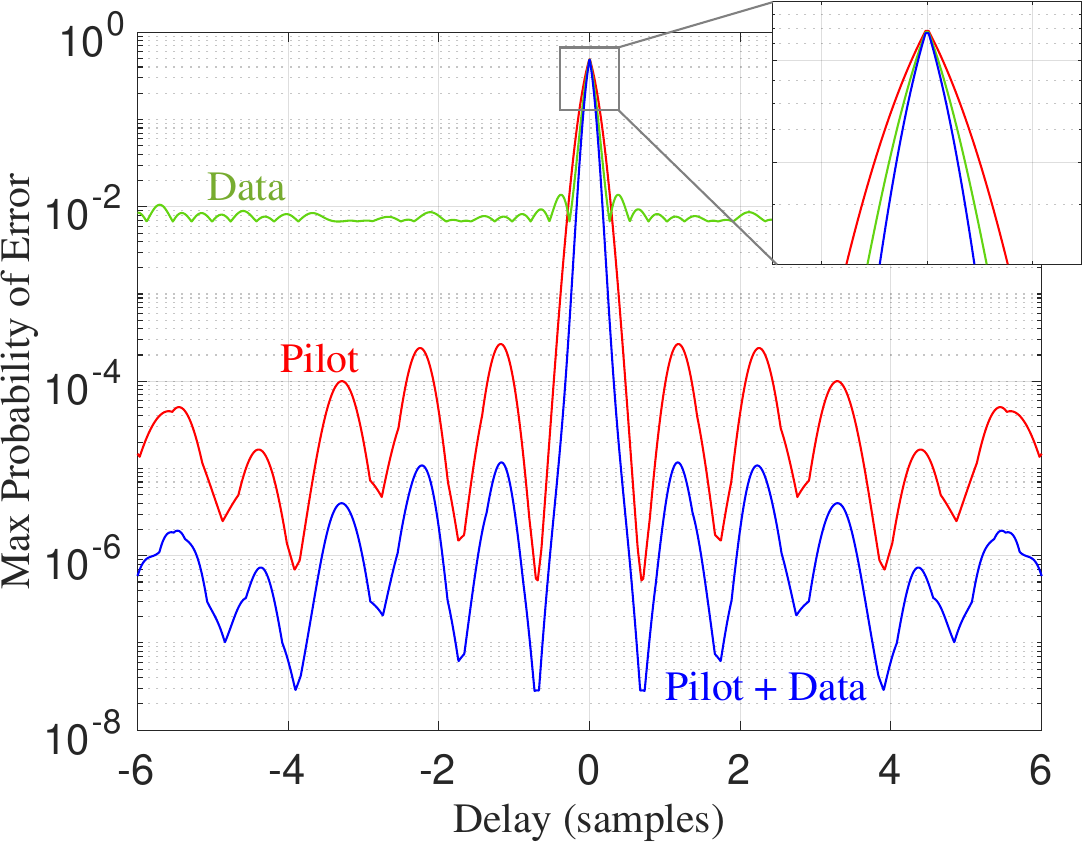}
	\caption{A plot of the maximin probability of error $\max_{\phi} \left[P_{\text{min}}(z,\phi)\right]$ against delay $z$ computed as part of the \ac{zzb} in (\ref{eq:unknown_data:zzb_vec}) for pilot-only, data-only, and pilot-plus-data estimation. Results are shown for the same signal in Fig.~\ref{fig:unknown_data:bound_comparison2} at \SI{5}{\decibel} \ac{snr}.}
	\label{fig:unknown_data:p_err}
\end{figure}

Fig.~\ref{fig:unknown_data:bound_comparison} compares the derived \ac{zzb} in (\ref{eq:unknown_data:zzb_vec}), the \ac{crlb} in (\ref{eq:unknown_data:bellili_crlb}), and the \ac{mcrlb} in (\ref{eq:unknown_data:mcrlb}) on \ac{toa} estimation against Monte-Carlo-simulation-based \acp{rmse} assuming the data-only \ac{ml} estimator in (\ref{eq:unknown_data:ml_d}). The \ac{toa} \acp{rmse} are scaled by the speed of light and plotted in units of meters. The \ac{ofdm} signal consists of $K=64$ subcarriers, $N_{\text{sym}}=1$ symbols, a subcarrier spacing of $\Delta_{\text{f}} = \SI{15}{\kilo\hertz}$, and an \textit{a priori} \ac{toa} duration of $\Ta = \SI{6.25}{\micro\second}$. All $64$ resource elements are allocated as data resources. Empirical \ac{toa} \acp{rmse} were estimated over $20000$ Monte Carlo iterations of random noise at each \ac{snr}. The grid search was conducted with intervals of $\Delta_{z} = \SI[parse-numbers = false]{1/8}{}$ sample and $\Delta_{\phi} = \SI{15}{\degree}$.

The \ac{mcrlb} is the loosest bound, only converging with the empirical \acp{rmse} at high \acp{snr} above \SI{10}{\decibel} for QPSK and \SI{17}{\decibel} for 16QAM. The \ac{crlb} in (\ref{eq:unknown_data:bellili_crlb}) remains tighter than the \ac{mcrlb} over a larger range of \acp{snr}, capturing the slight deviation from the \ac{mcrlb} above \SI{7}{\decibel} \ac{snr} for QPSK and \SI{14}{\decibel} \ac{snr} for 16QAM. Below these \acp{snr}, the empirical \ac{toa} errors experience the low-\ac{snr} thresholding effect and \acp{rmse} increase suddenly, a phenomenon not captured by the \ac{crlb}. The \ac{zzb} provides a much tighter bound in this thresholding regime than the \ac{mcrlb} and \ac{crlb}. Asymptotically as \ac{snr} decreases, the \ac{zzb} and empirical \acp{rmse} converge to different values since the empirical \ac{toa} estimator is \ac{ml}, not \ac{mmse}. Accordingly, the \ac{zzb} \ac{rmse} converges to $\sqrt{\Ta/12}\; \SI{}{\second}$, the standard deviation of a uniform distribution with duration $\Ta\; \SI{}{\second}$ \cite{chazan1975improved}.

Fig.~\ref{fig:unknown_data:bound_comparison2} provides a different perspective and compares the empirical \acp{rmse} against their \acp{zzb} for the four types of estimators: pilot-only \ac{ml}, data-only \ac{ml}, pilot-plus-data \ac{ml}, and \ac{dd}. The \ac{ofdm} signal is parameterized identically to the signal in Fig.~\ref{fig:unknown_data:bound_comparison} but is additionally allocated with a sparse placement of $8$ pilot resources. The pilot resource placement is optimized to minimize \ac{toa} error at \SI{0}{\decibel} \ac{snr} using the integer-optimization routines in \cite{graff2024ziv}. Power is allocated equally across all resources.

Below \SI{8}{\decibel} \ac{snr}, the pilot-only estimator reduces error compared to the data-only estimator. Above this \ac{snr}, however, the data-only estimator reduces \ac{toa} errors significantly over the pilot-only estimator, ultimately achieving an \ac{rmse} of  \SI{3.0}{\meter} compared to the pilot-only estimator's \ac{rmse} of \SI{5.8}{\meter} at \SI{15}{\decibel} \ac{snr}. The pilot-plus-data estimator achieves the lowest \ac{rmse} across all \acp{snr}. This is most notable in the thresholding regime, with the pilot-plus-data estimator achieving an \ac{rmse} of \SI{11.8}{\meter} compared to pilot-only \ac{rmse} of \SI{54.5}{\meter} and data-only \ac{rmse} of \SI{338.2}{\meter} at \SI{5}{\decibel} \ac{snr}. Meanwhile, the \ac{dd} estimator is only capable of improving upon the pilot-only estimator at and above \SI{6}{\decibel} \ac{snr}, after which it plateaus near the same \ac{rmse} as the pilot-plus-data estimator.

Fig.~\ref{fig:unknown_data:p_err} provides insight into how the different types of estimation affect the probability of error in (\ref{eq:unknown_data:p_err}), thereby changing the \ac{zzb} in (\ref{eq:unknown_data:zzb_vec}). As seen in Fig.~\ref{fig:unknown_data:bound_comparison2}, all three \acp{zzb} at \SI{5}{\decibel} \ac{snr} are experiencing the low-\ac{snr} thresholding effect to varying degrees. At this \ac{snr}, Fig.~\ref{fig:unknown_data:p_err} shows that the pilot-only probability of error exhibits high sidelobes and a wide peak near the true delay, resulting in the large \ac{rmse} in Fig.~\ref{fig:unknown_data:bound_comparison2}. Meanwhile, the data-only probability of error exhibits a sharper peak but higher sidelobes that remain relatively flat across all delays. This elevated sidelobe presence significantly increases the likelihood of \ac{toa} estimates occurring outside of the mainlobe. As a result, data-only estimation has the highest \ac{rmse} in Fig.~\ref{fig:unknown_data:bound_comparison2} at \SI{5}{\decibel} \ac{snr} despite the sharpened mainlobe. In contrast, the pilot-plus-data probability of error exhibits both the sharpest mainlobe peak and the lowest sidelobe probability, allowing pilot-plus-data estimation to mitigate the low-\ac{snr} thresholding effect significantly and achieve a lower \ac{rmse} than the pilot-only and data-only estimators.

\subsection{Sparse Resource Optimization}
\label{sec:unknown_data:results_prs}

Now consider the problem of allocating \acp{prs} in an \ac{ofdm} signal designed for both positioning and communications. To minimize the reduction in data rate, the \acp{prs} will be allocated sparsely throughout the bandwidth of the signal. Consider an \ac{ofdm} signal consisting of $K=240$ subcarriers, $N_{\text{sym}}=4$ symbols, a subcarrier spacing of $\Delta_{\text{f}} = \SI{240}{\kilo\hertz}$, and an \textit{a priori} \ac{toa} duration of $\Ta = \SI{156.25}{\nano\second}$. Assume the subcarriers are divided into $20$ resource blocks of 12 subcarriers each, where each resource block is restricted to containing either a \ac{prs} block or a data block filled with QPSK data resources. Letting $N_{\text{PRS}}$ denote the number of resource blocks that are allocated with a \ac{prs} block, the allocation problem is to determine the best placement of these $N_{\text{PRS}}$ \ac{prs} blocks among the $20$ available resource blocks.

\begin{figure}[h]
	\centering
	\includegraphics[width=0.9\linewidth]{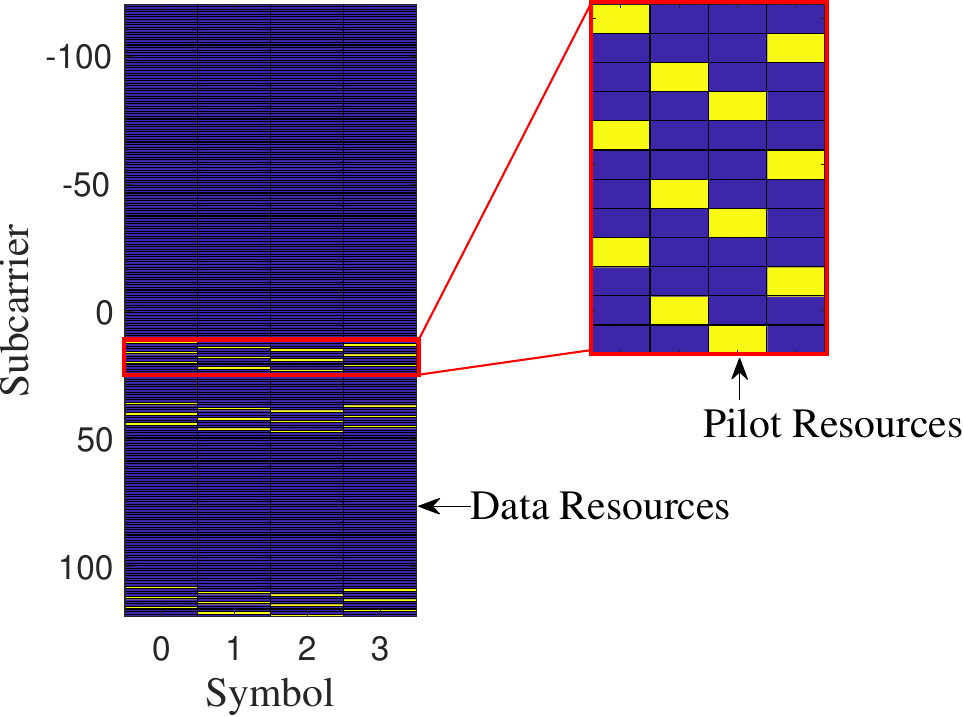}
	\caption{A visualization of the optimized \ac{ofdm} resource allocation for $N_{\text{PRS}}=3$ at \SI{0}{\decibel} \ac{snr}.}
	\label{fig:unknown_data:prs_vis}
\end{figure}

Each \ac{prs} block has been arbitrarily chosen to consist of pilot resources placed in a size $4$ comb pattern, similarly to the \ac{prs} in 5G NR, which is visualized in Fig.~\ref{fig:unknown_data:prs_vis}. All non-pilot resource elements in each \ac{prs} block are allocated as data resources.

\begin{figure}[h]
	\centering
	\includegraphics[width=0.9\linewidth]{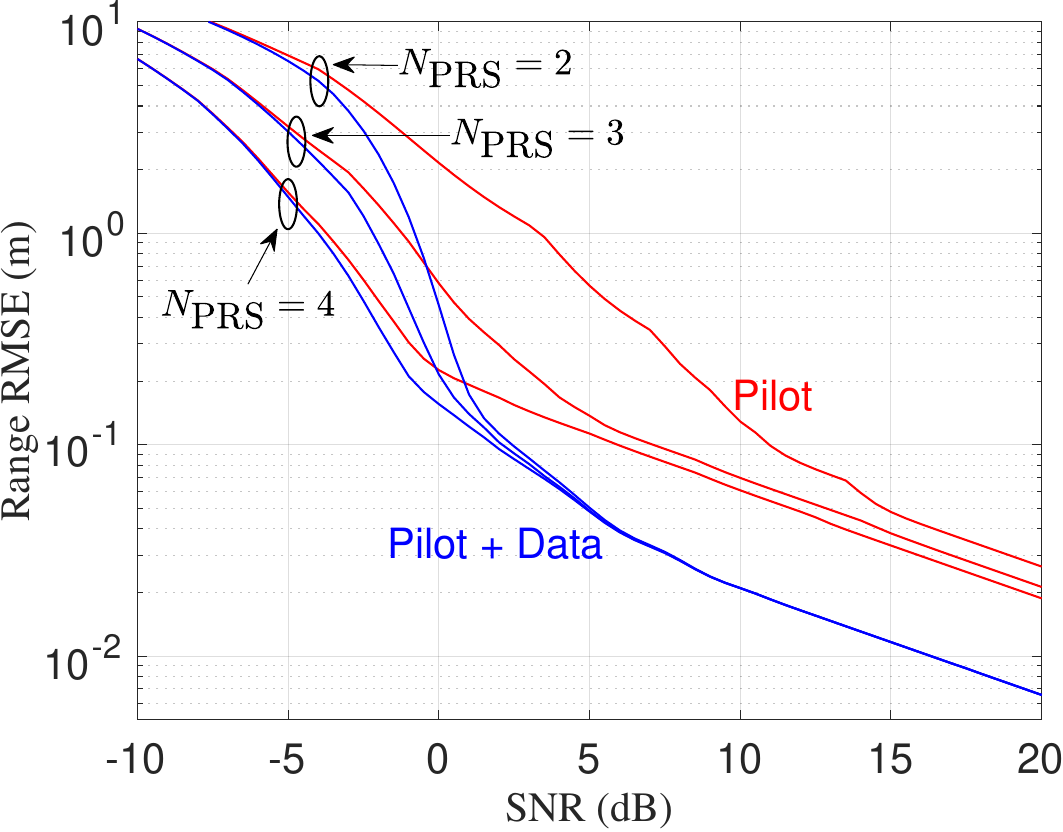}
	\caption{The \ac{zzb} on \ac{toa} \ac{rmse} for the PRS-optimized resource allocations. Results are shown for $N_{\text{PRS}} \in \{2,3,4\}$.}
	\label{fig:unknown_data:prs_optim_zzb}
\end{figure}

The pilot-only \ac{zzb} was evaluated for all permutations of \ac{prs} allocation for $N_{\text{PRS}} \in \{2,3,4\}$ at each \ac{snr}. The pilot-optimal allocation at each \ac{snr} was then chosen as the allocation that minimized the \ac{zzb}. The pilot-plus-data \ac{zzb} was then evaluated on the pilot-optimal allocations. Fig.~\ref{fig:unknown_data:prs_optim_zzb} plots both the pilot-only and pilot-plus-data \acp{zzb} of the pilot-optimal allocations against \ac{snr}. Increasing the number of \ac{prs} resource blocks reduces the \ac{zzb} for both pilot-only and pilot-plus-data estimation, yielding the greatest improvement at lower \acp{snr} where the additional pilot resources can mitigate the low-\ac{snr} thresholding effect. However, this improvement becomes negligible for the pilot-plus-data \acp{zzb} above approximately \SI{6}{\decibel} \ac{snr} where the bounds converge. In this high \ac{snr} regime, the pilot-plus-data \acp{zzb} show significant reductions in error compared to the pilot-only \acp{zzb}. At \SI{10}{\decibel} \ac{snr}, all three pilot-plus-data \acp{zzb} have a \ac{rmse} of \SI{2.1}{\centi\meter} compared to \acp{rmse} of \SI{6.1}{\centi\meter}, \SI{7.0}{\centi\meter}, and \SI{12.9}{\centi\meter} for the pilot-only \acp{zzb}. Into low \acp{snr}, the pilot-plus-data \acp{zzb} still show notable improvements over their respective pilot-only \acp{zzb} even though the \ac{prs} allocations are optimized for pilot-only estimation at every \ac{snr}.

\subsection{LEO Satellite Positioning}
\label{sec:unknown_data:results_leo}

Positioning with \ac{leo} satellite downlink signals is a particularly apt application for the \ac{ml} and \ac{dd} \ac{toa} estimators discussed in Section~\ref{sec:unknown_data:ml_estimation}. \ac{leo} channels can span wide bandwidths, enabling highly-accurate \ac{toa} estimates and therefore accurate user positioning services. Furthermore, \ac{leo} channels experience minimal fading especially when combined with highly-directional phased arrays at both the transmitter and receiver, resulting in flat channel responses across the wide signal bandwidths. Finally, \ac{leo} satellites provide communication services to large cells which are likely to contain many network users, increasing the amount of downlink data resources that will need to be allocated. If the \ac{leo} satellites transmit in bursts to manage power consumption \cite{iannucci2022fusedLeo}, downlink resources are likely to be fully allocated to maximize throughput. One example is the downlink Starlink signal which consists of frames with fully allocated data \cite{humphreys2023starlinkSignalStructure}. In such a fully allocated burst, the allocation of \acp{prs} comes at the cost of decreased throughput. Therefore, it may be beneficial to allocate fewer resources to positioning services and instead have receivers exploit the unknown data resources to obtain accurate positioning. In this section, the positioning performance of the \ac{ml} and \ac{dd} \ac{toa} estimators is evaluated on a simulated \ac{leo} downlink channel.

\subsubsection*{Setup}

\begin{figure}[h]
	\centering
	\includegraphics[width=0.9\linewidth]{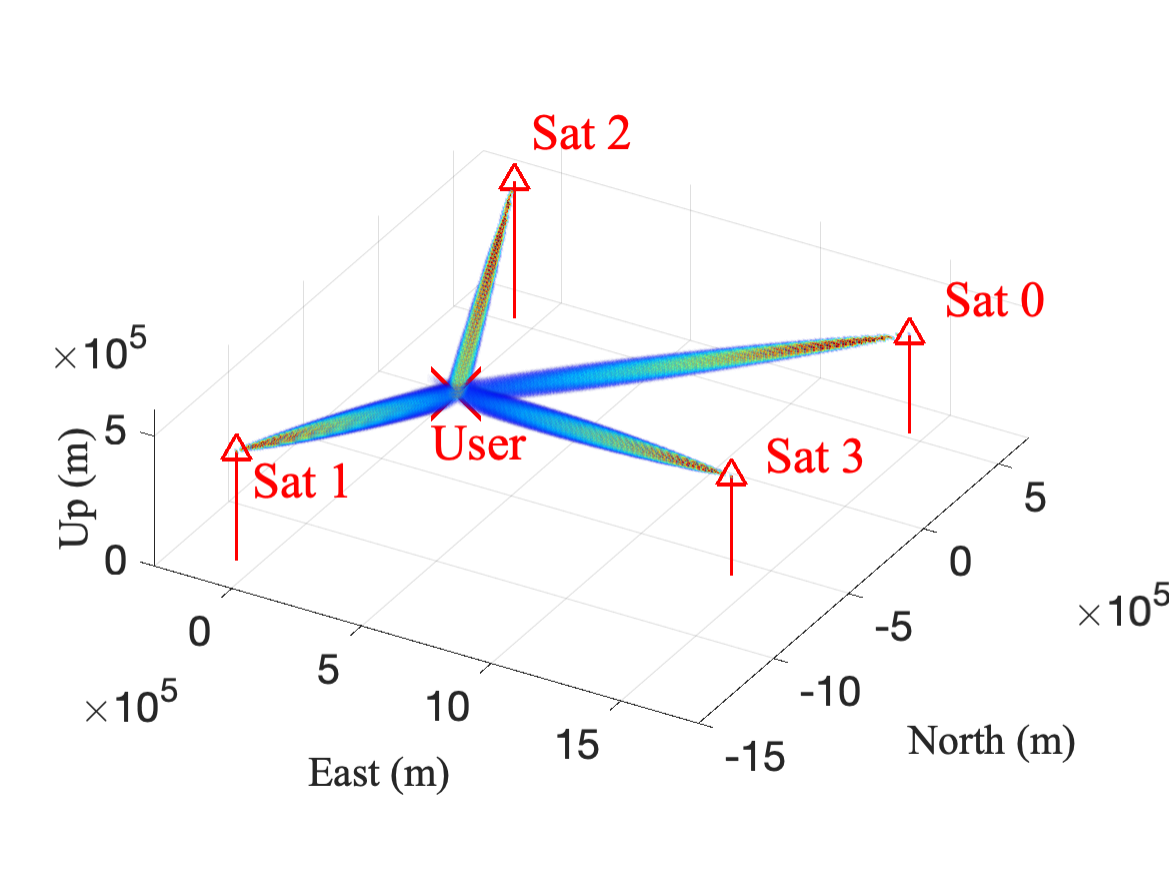}
	\caption{A visualization of the four satellites in the simulated LEO constellation and each satellite's downlink beam.}
	\label{fig:unknown_data:sat_vis}
\end{figure}

The simulated environment consists of four \ac{leo} satellites and one receiver. The receiver's cell receives downlink service from each satellite in a time-duplexed manner with the interval between downlink bursts from each satellite denoted as $T_{\text{burst}}$. During each downlink burst, the servicing satellite directs its beam to the center of the cell and pre-compensates for the expected Doppler shift experienced by a stationary user at the center of the cell. For simplicity, the receiver is stationary and located at the center of its cell. The receiver also has perfect knowledge of each satellite's relative position and applies conventional beamforming weights to its phased array elements to direct its beam in the known direction of the servicing satellite. Fig.~\ref{fig:unknown_data:sat_vis} shows a graphical depiction of the simulation setup and the downlink beams.

Each burst consists of $N_{\text{sym}} = 4$ \ac{ofdm} symbols. The \ac{ofdm} signal parameters and simulation parameters are listed in Table~\ref{tab:unknown_data:sat_params}. Simulated frequency-selective, time-varying channels are generated using QuaDRiGa \cite{burkhardt2014quadriga}. Two channel models are considered: \textit{QuaDRiGa\_NTN\_DenseUrban\_LOS} \cite{jaeckel20225g} and \textit{5G-ALLSTAR\_DenseUrban\_LOS} \cite{cassiau2020satellite}.

For each channel model, $1000$ channel realizations were generated. For each channel realization, $1000$ realizations of \ac{awgn} were generated. Pseudorange estimates were obtained for each channel and \ac{awgn} realization using each of the estimators Additionally, the \ac{zzb} was computed for each channel realization.

Let $\bm{r}_i$ be the \ac{enu} coordinates of satellite $i$ for $i \in \{0,1,2,3\}$, $\bm{r}$ be the \ac{enu} coordinates of the receiver, and $\delta_{t}$ be the clock offset between the receiver and the satellite constellation. Define $c$ as the speed of light, $\bm{\epsilon}_{\rho}$ as pseudorange estimation error, and $\bm{\theta}_{\rho} \triangleq [\bm{r}^T,\; c\delta_{t}]^T$. Additionally define
\begin{align}
	h_{i}(\bm{\theta}_{\rho}) \triangleq c||\bm{r}_{i} - \bm{r}|| + c\delta_{t},
\end{align}
which is expressed in vector form as $\bm{h}(\bm{\theta}_{\rho}) \triangleq [h_{0}(\bm{\theta}_{\rho}),\; h_{1}(\bm{\theta}_{\rho}),\; h_{2}(\bm{\theta}_{\rho}),\; h_{3}(\bm{\theta}_{\rho})]^T$.
Then the vectorized pseudorange measurement equation for all satellites becomes
\begin{align}
	\bm{\rho}= \bm{h}(\bm{\theta}_{\rho}) + \bm{\epsilon}_{\rho}.
\end{align}
Letting $\bm{\Sigma}_{\bm{\rho}}$ be the covariance of the pseudorange error $\bm{\epsilon}_{\rho}$, a positioning solution can then be obtained by solving the weighted nonlinear least-squares problem:
\begin{align}
	\hat{\bm{\theta}}_{\rho} = \argmin_{\bm{\theta}_{\rho}} \bigl|\bigl| \bm{\rho} - \bm{h}(\bm{\theta}_{\rho}) \bigr|\bigr|_{\bm{\Sigma}_{\bm{\rho}}^{-1}}^2.
\end{align}
The error covariance of this estimate can be approximated by linearizing the pseudorange residuals at the true values of $\bm{\theta}_{\rho}$. Defining the matrix
\begin{align}
	\bm{A} \triangleq 
	\begin{bmatrix}
		\sfrac{\bm{r}_{0}-\bm{r}}{||\bm{r}_{0}-\bm{r}||^2} & \sfrac{\bm{r}_{1}-\bm{r}}{||\bm{r}_{1}-\bm{r}||^2} & \sfrac{\bm{r}_{2}-\bm{r}}{||\bm{r}_{2}-\bm{r}||^2} & \sfrac{\bm{r}_{3}-\bm{r}}{||\bm{r}_{3}-\bm{r}||^2} \\
		1 & 1 & 1 & 1
	\end{bmatrix}^{T},
\end{align}
%
the error covariance of $\hat{\bm{\theta}}_{\rho}$ can then be described as $\bm{Q} \triangleq \left(\bm{A}^{T} \bm{\Sigma}_{\bm{\rho}}^{-1}\bm{A}\right)^{-1}$. The diagonal elements of this error covariance are defined as $[\sigma_{x}^2,\;\sigma_{y}^2,\;\sigma_{z}^2,\;\sigma_{c\delta{t}}^2]^T \triangleq \text{diag}(\bm{Q})$.

Since each pseudorange is obtained independently, $\bm{\Sigma}_{\bm{\rho}}$ is a diagonal matrix consisting of elements $\sigma_{\rho_0}^2$, $\sigma_{\rho_1}^2$, $\sigma_{\rho_2}^2$, and $\sigma_{\rho_3}^2$. These variances are obtained either from the \ac{zzb} or from the empirical error variance of the \ac{toa} estimates over \ac{awgn} realizations for a single channel realization. It is important to note that the distribution of the \ac{toa} errors is unknown and not guaranteed to be Gaussian. However, the \acp{zzb} and empirical error variances can be compared through this linearized model.

\begin{table}[t]
	\centering
	\caption{LEO Simulation Parameters}
	\begin{tabular}[c]{ll}
		\toprule
		\multicolumn{2}{c}{OFDM Parameters} \\
		\midrule
		$K$ & $240$ subcarriers \\
		$\Delta_{\text{f}}$ (Subcarrier Spacing) & \SI{240}{\kilo\hertz} \\
		Symbol Duration & \SI{4.167}{\micro\second} \\
		Cyclic Prefix Duration &  \SI{0.521}{\micro\second} \\
		Bandwidth & \SI{57.60}{\mega\hertz} \\
		Symbol Constellation & QPSK \\
		\toprule
		\multicolumn{2}{c}{Simulation Parameters} \\
		\midrule
		$\fc$ (Carrier Frequency) & \SI{10.7}{\giga\hertz} \\
		Polarization & LHCP @ RX \& TX \\
		TX Gain & \SI{34}{\decibel} \\
		TX Beamwidth & \SI{3.67}{\degree} \\
		RX Array & $32\times32$ URA $\left(\lambda/2\;\text{spacing}\right)$ \\
		RX Gain & \SI{30}{\decibel} \\
		RX Beamwidth & \SI{3.58}{\degree} \\
		EIRP & \SI[per-mode=symbol]{-15}{\decibel\watt\per4\kilo\hertz} \\
		$\sigma^2$ (Resource Noise Power)& $-173.8 + 10\log_{10}(\Delta_{\text{f}})$ \SI{}{\deci\belm}\\
		$\Ta$ (\textit{a priori} \ac{toa} Duration) & \SI{156.25}{\nano\second} \\
		$T_{\text{burst}}$ (Burst Interval) & \SI{1}{\milli\second} \\
		Satellite Altitude & \SI{550}{\kilo\meter} \\
		Satellite Constellation & Walker-Delta \SI{53}{\degree}:1584/22/39 \\
		Elevation Mask & \SI{30}{\degree} \\
		Channel Models & \textit{QuaDRiGa\_NTN\_DenseUrban\_LOS} \\
		& \textit{5G-ALLSTAR\_DenseUrban\_LOS} \\
		\bottomrule
	\end{tabular}
	\label{tab:unknown_data:sat_params}
\end{table}

\subsubsection*{LEO Results}


\begin{figure}[h]
	\centering
	\includegraphics[width=0.9\linewidth]{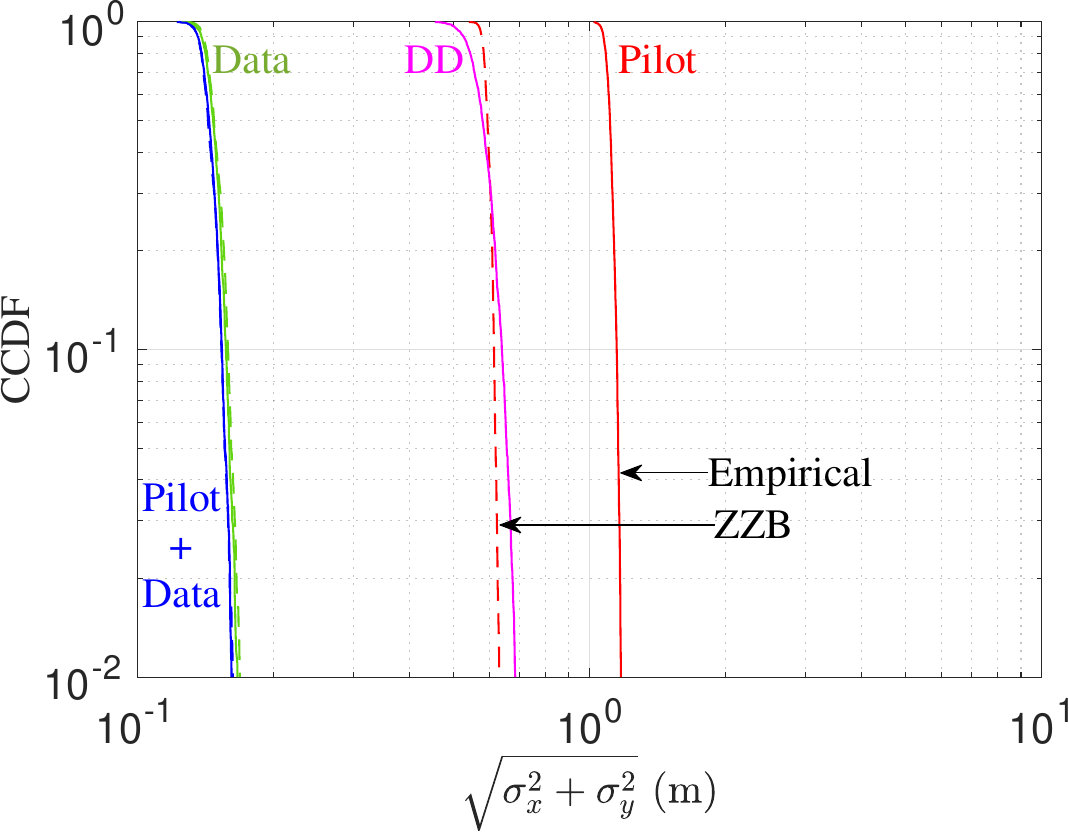}
	\caption{The \acp{ccdf} of horizontal \ac{rmse} over channel realizations using the QuaDRiGa channel model, comparing the pilot-only, data-only, pilot-plus-data, and \ac{dd} estimators. Results are shown for both the \ac{zzb} and empirical data.}
	\label{fig:unknown_data:horiz_ccdf_quad}
\end{figure}

\begin{figure}[h]
	\centering
	\includegraphics[width=0.9\linewidth]{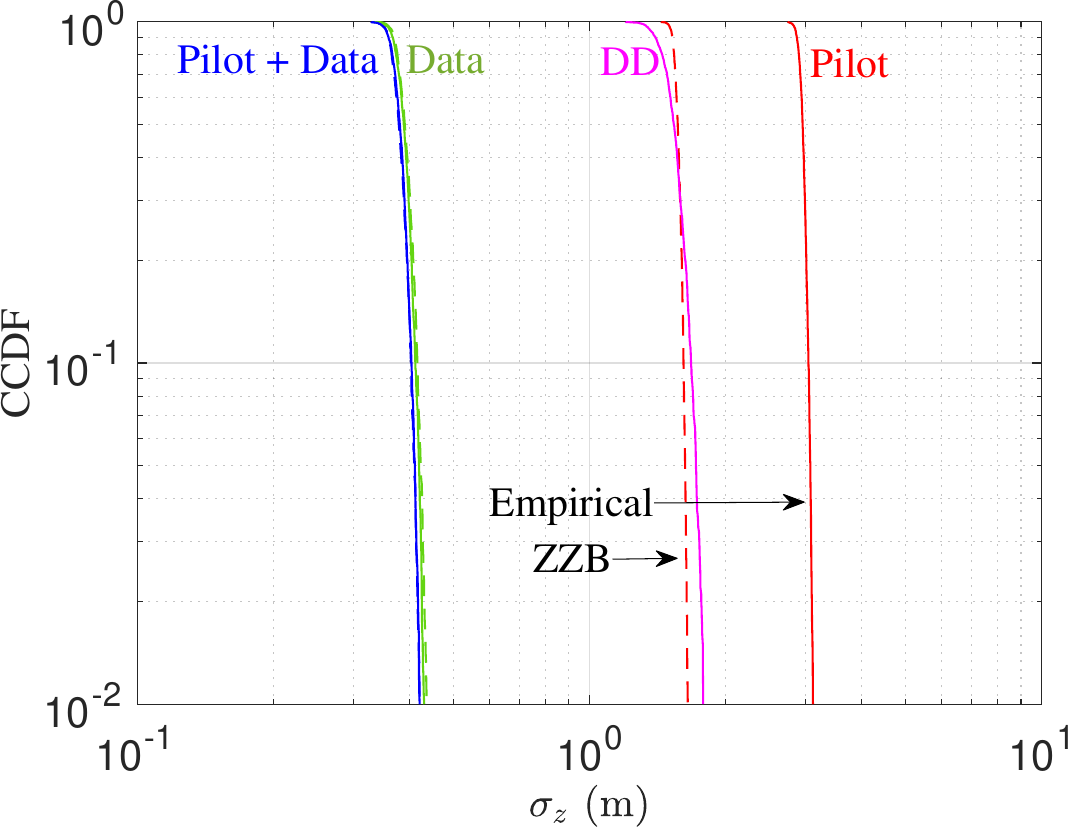}
	\caption{As Fig.~\ref{fig:unknown_data:horiz_ccdf_quad} but for vertical \ac{rmse}.}
	\label{fig:unknown_data:vert_ccdf_quad}
\end{figure}

\begin{figure}[h]
	\centering
	\includegraphics[width=0.9\linewidth]{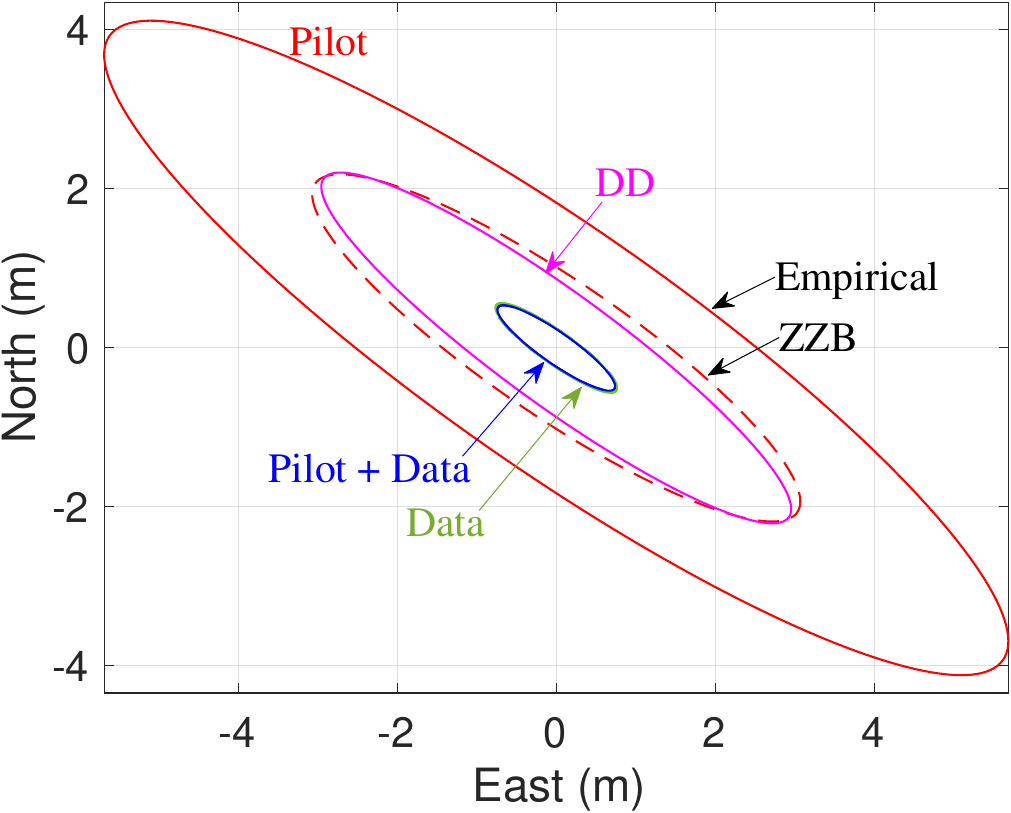}
	\caption{The \SI{95}{\percent} horizontal positioning error ellipses for the QuaDRiGa channel model. Results are obtained using the Chebyshev inequality with both the \ac{zzb} and empirical \acp{rmse} over all channel realizations.}
	\label{fig:unknown_data:err_ellipse_quad}
\end{figure}

Fig.~\ref{fig:unknown_data:horiz_ccdf_quad} and Fig.~\ref{fig:unknown_data:vert_ccdf_quad} plot the \ac{ccdf} over channel realizations of the horizontal \acp{rmse} and vertical \acp{rmse} using the \textit{QuaDRiGa\_NTN\_DenseUrban\_LOS} channel model. Both the \acp{zzb} and empirical \acp{rmse} are depicted for each of the estimators. The pilot-only estimator results in the greatest errors, having a $90$th percentile horizontal \ac{rmse} of \SI{1.15}{\meter} and vertical \ac{rmse} of \SI{3.05}{\meter}. The decision-directed estimator yields moderate improvements over the pilot-only estimator, having a $90$th percentile horizontal \ac{rmse} of \SI{0.64}{\meter} and vertical \ac{rmse} of \SI{1.68}{\meter}. Accuracy is improved significantly with the data-only estimator, having a $90$th percentile horizontal \ac{rmse} of \SI{0.16}{\meter} and vertical \ac{rmse} of \SI{0.41}{\meter}. Finally, the pilot-plus-data estimator achieves the best accuracy, having a $90$th percentile horizontal \ac{rmse} of \SI{0.15}{\meter} and vertical \ac{rmse} of \SI{0.40}{\meter}.

Fig.~\ref{fig:unknown_data:err_ellipse_quad} provides an alternative depiction of these results, visualizing the \SI{95}{\percent} error ellipses using both the \ac{zzb} and empirical \acp{rmse} over all channel realizations. Since the error distribution is not guaranteed to be Gaussian, the \SI{95}{\percent} error ellipses are computed using the multivariate Chebyshev inequality \cite{marshall1960multivariate}, resulting in a 2D ellipse corresponding to $6.32$ standard deviations. The Chebyshev inequality allows confidence intervals to be constructed for any arbitrary distribution with a finite variance. This depiction highlights how the data-only and pilot-plus-data estimators reduce positioning error significantly compared to the pilot-only estimator and even the \ac{dd} estimator. The data-only and pilot-plus-data ellipses have a semi-major axis of approximately \SI{0.9}{\meter} compared to the \ac{dd} ellipse's \SI{3.6}{\meter} and pilot-only ellipse's \SI{6.9}{\meter}. The \ac{zzb} ellipses are nearly coincident with the empirical ellipses for data-only and pilot-plus-data estimation, while the \ac{zzb} ellipse for pilot-only estimation has a gap compared to the empirical ellipse and a semi-major axis of \SI{3.7}{\meter}.

The pilot-plus-data and data-only estimators are capable of outperforming the pilot-only estimator by exploiting significantly more resources in the signal. Compared to the DD estimator, these estimators are not negatively impacted by errors in the hard decoding process. As a result, the DD estimator requires much higher \acp{snr} to approach the performance of the pilot-plus-data and data-only estimators.

\begin{figure}[h]
	\centering
	\includegraphics[width=0.9\linewidth]{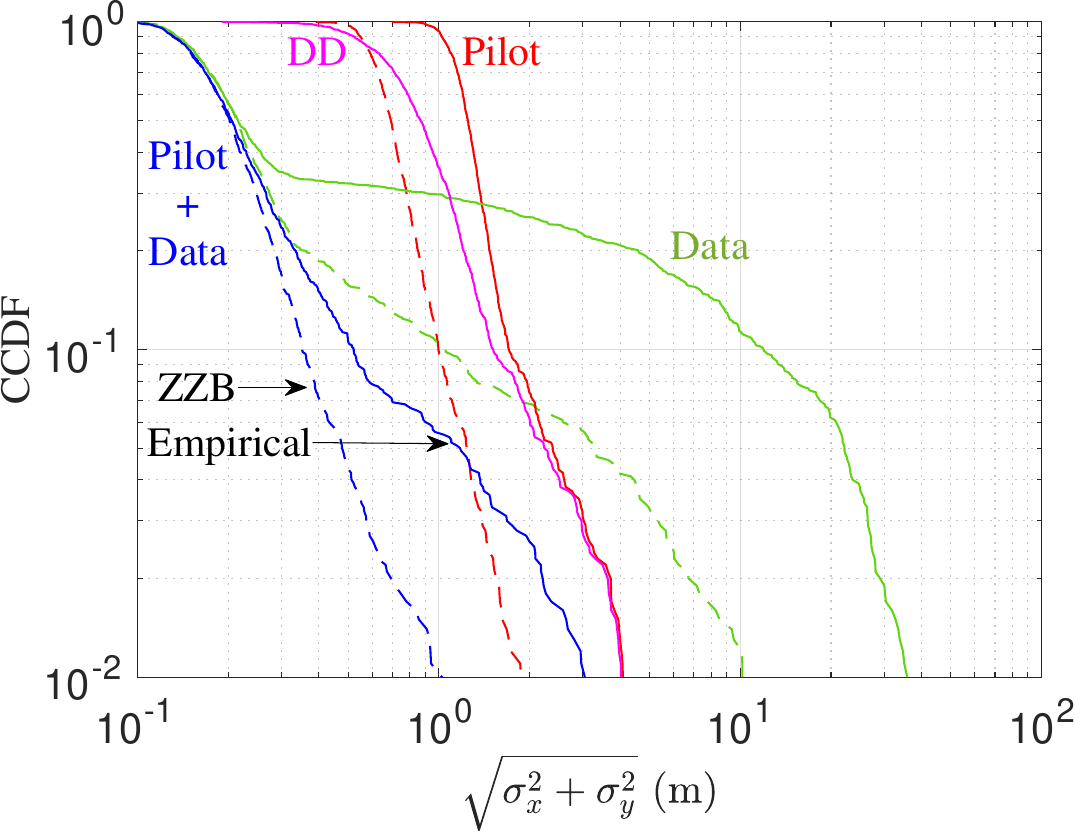}
	\caption{The \acp{ccdf} over channel realizations of horizontal \ac{rmse} using the 5G-ALLSTAR channel model, comparing the pilot-only, data-only, pilot-plus-data, and \ac{dd} estimators. Results are shown for both the \ac{zzb} and empirical data.}
	\label{fig:unknown_data:horiz_ccdf_allstar}
\end{figure}

\begin{figure}[h]
	\centering
	\includegraphics[width=0.9\linewidth]{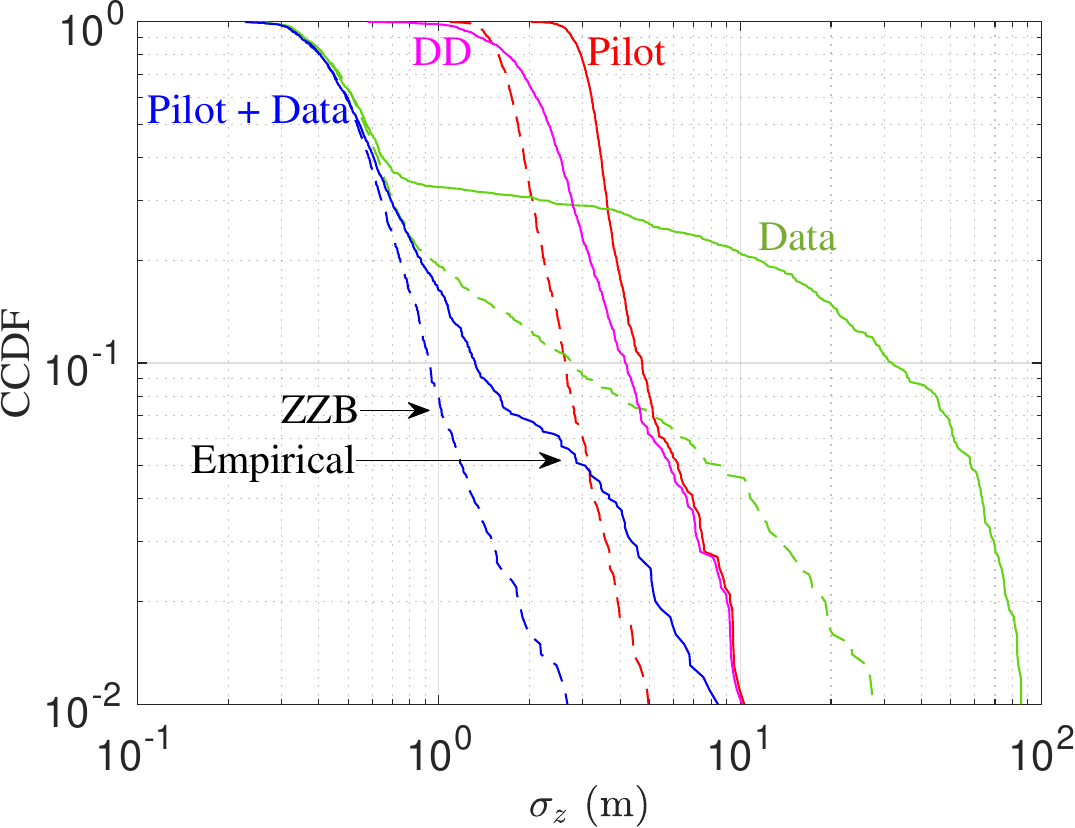}
	\caption{As Fig.~\ref{fig:unknown_data:horiz_ccdf_allstar} but for vertical \ac{rmse}.}
	\label{fig:unknown_data:vert_ccdf_allstar}
\end{figure}

\begin{figure}[h]
	\centering
	\includegraphics[width=0.9\linewidth]{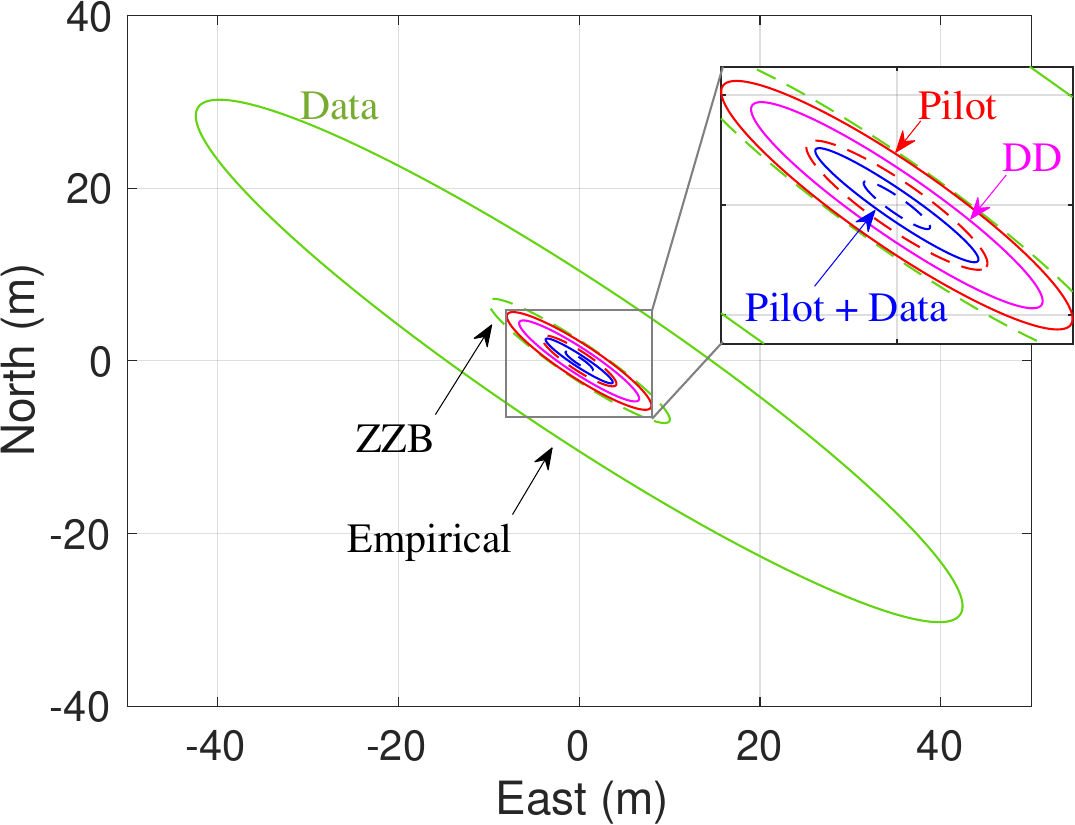}
	\caption{The \SI{95}{\percent} horizontal positioning error ellipses for the 5G-ALLSTAR channel model. Results are obtained using the Chebyshev inequality with both the \ac{zzb} and empirical \acp{rmse} over all channel realizations.}
	\label{fig:unknown_data:err_ellipse_allstar}
\end{figure}

Figs.~\ref{fig:unknown_data:horiz_ccdf_allstar} and \ref{fig:unknown_data:vert_ccdf_allstar} plot the \ac{ccdf} over channel realizations of the horizontal \acp{rmse} and vertical \acp{rmse} using the \textit{5G-ALLSTAR\_DenseUrban\_LOS} channel model, similar to Figs.~\ref{fig:unknown_data:horiz_ccdf_quad} and \ref{fig:unknown_data:vert_ccdf_quad}. Different patterns emerge with this channel model, as greater fluctuations in \ac{snr} are simulated. Similar to the results in Figs.~\ref{fig:unknown_data:horiz_ccdf_quad} and \ref{fig:unknown_data:vert_ccdf_quad}, the pilot-plus-data estimator achieves the greatest accuracy, having a $90$th percentile horizontal \ac{rmse} of \SI{0.52}{\meter} and vertical \ac{rmse} of \SI{1.32}{\meter}. However, the data-only estimator only outperforms the DD and pilot-only estimators in approximately \SI{70}{\percent} of channel realizations. In the remaining \SI{30}{\percent} of channel realizations, the low \acp{snr} cause the data-only estimator to enter its thresholding regime, resulting in \acp{rmse} surpassing those achieved using only pilots. The data-only estimator has a $90$th percentile horizontal \ac{rmse} of \SI{11.96}{\meter} and vertical \ac{rmse} of \SI{32.00}{\meter}. Meanwhile, the \ac{dd} estimator exhibits less improvement over the pilot-only estimator in this channel model, having a $90$th percentile horizontal \ac{rmse} of \SI{1.52}{\meter} and vertical \ac{rmse} of \SI{4.16}{\meter} compared to the pilot-only estimator's $90$th percentile horizontal \ac{rmse} of \SI{1.71}{\meter} and vertical \ac{rmse} of \SI{4.73}{\meter}.


Fig.~\ref{fig:unknown_data:err_ellipse_allstar} visualizes the \SI{95}{\percent} error ellipses for the 5G-ALLSTAR channel model, similar to Fig.~\ref{fig:unknown_data:err_ellipse_quad}. As with the results in Fig.~\ref{fig:unknown_data:horiz_ccdf_allstar} and Fig.~\ref{fig:unknown_data:vert_ccdf_allstar}, the data-only estimator exhibits significant errors due to the poor channel conditions. Meanwhile, the pilot-plus-data estimator reduces postioning error compared to both the pilot-only and \ac{dd} estimators. The pilot-plus-data ellipse has a semi-major axis of approximately \SI{4.5}{\meter} compared to the \ac{dd} ellipse's \SI{8.0}{\meter}, pilot-only ellipse's \SI{9.6}{\meter}, and data-only ellipse's \SI{52}{\meter}.

The large \ac{snr} fluctuations in the \textit{5G-ALLSTAR\_DenseUrban\_LOS} channel model resulted in the estimators entering their low-\ac{snr} thresholding regimes. The pilot-plus-data estimator provides robustness against these thresholding effects at low \ac{snr} while simultaneously maximizing accuracy in high \ac{snr}. In comparison to Fig.~\ref{fig:unknown_data:horiz_ccdf_quad} and Fig.~\ref{fig:unknown_data:vert_ccdf_quad}, the \acp{zzb} are much looser due to the bound not being as tight in the low-\ac{snr} thresholding regime as in the high-\ac{snr} regime.

\section{Conclusions}
\label{sec:unknown_data:conclusion}

This paper has derived a novel \ac{zzb} on \ac{toa} estimation with \ac{ofdm} signals containing unknown data resources. This \ac{zzb} serves as a lower bound to both \ac{ml} and \ac{dd} estimators that can exploit unknown data resources to improve estimation accuracy. The \ac{zzb} was shown to be tighter to empirical errors than the \ac{crlb} and \ac{mcrlb} derived in prior work, making it a useful criterion for evaluating different \ac{ofdm} resource allocations for \ac{toa} estimation. Comparisons were then made between four different types of estimators: pilot-only \ac{ml}, data-only \ac{ml}, pilot-plus-data \ac{ml}, and \ac{dd}, demonstrating that the pilot-plus-data estimator can significantly improve \ac{toa} estimation accuracy. The \ac{zzb} was then used to evaluate different allocations of \acp{prs} within a wideband \ac{ofdm} signal, which can guide resource allocations that minimize overhead while still achieving \ac{toa} accuracy requirements. Finally, the \acp{zzb} and \ac{toa} estimators were evaluated on simulated \ac{leo} channels, quantifying the distribution of positioning \acp{rmse} across channel realizations. These results highlight the potential for \ac{ofdm} networks to significantly improve positioning performance while still prioritizing data throughput.
\section*{Acknowledgments}
This work was supported by the U.S. Department of Transportation under Grant
69A3552348327 for the CARMEN+ University Transportation Center, by the
U.S. Space Force under an STTR contract with Coherent Technical Services, Inc.,
and by affiliates of the 6G@UT center within the Wireless Networking and
Communications Group at The University of Texas at Austin.


\bibliographystyle{IEEEtran} 
\bibliography{pangea}

\begin{thebibliography}{10}
\providecommand{\url}[1]{#1}
\csname url@samestyle\endcsname
\providecommand{\newblock}{\relax}
\providecommand{\bibinfo}[2]{#2}
\providecommand{\BIBentrySTDinterwordspacing}{\spaceskip=0pt\relax}
\providecommand{\BIBentryALTinterwordstretchfactor}{4}
\providecommand{\BIBentryALTinterwordspacing}{\spaceskip=\fontdimen2\font plus
\BIBentryALTinterwordstretchfactor\fontdimen3\font minus
  \fontdimen4\font\relax}
\providecommand{\BIBforeignlanguage}[2]{{%
\expandafter\ifx\csname l@#1\endcsname\relax
\typeout{** WARNING: IEEEtran.bst: No hyphenation pattern has been}%
\typeout{** loaded for the language `#1'. Using the pattern for}%
\typeout{** the default language instead.}%
\else
\language=\csname l@#1\endcsname
\fi
#2}}
\providecommand{\BIBdecl}{\relax}
\BIBdecl

\bibitem{dureppagari2023ntn}
H.~K. Dureppagari, C.~Saha, H.~S. Dhillon, and R.~M. Buehrer, ``{NTN}-based
  {6G} localization: Vision, role of {LEOs}, and open problems,'' \emph{IEEE
  Wireless Communications}, vol.~30, no.~6, pp. 44--51, 2023.

\bibitem{iannucci2022fusedLeo}
P.~A. Iannucci and T.~E. Humphreys, ``Fused low-earth-orbit {GNSS},''
  \emph{IEEE Transactions on Aerospace and Electronic Systems}, pp. 1--1, 2022.

\bibitem{lin2021path}
X.~Lin, S.~Cioni, G.~Charbit, N.~Chuberre, S.~Hellsten, and J.-F. Boutillon,
  ``On the path to {6G}: Embracing the next wave of low {Earth} orbit satellite
  access,'' \emph{IEEE Communications Magazine}, vol.~59, no.~12, pp. 36--42,
  2021.

\bibitem{dwivedi2021positioning}
S.~Dwivedi, R.~Shreevastav, F.~Munier, J.~Nygren, I.~Siomina, Y.~Lyazidi,
  D.~Shrestha, G.~Lindmark, P.~Ernstr{\"o}m, E.~Stare \emph{et~al.},
  ``Positioning in 5{G} networks,'' \emph{IEEE Communications Magazine},
  vol.~59, no.~11, pp. 38--44, 2021.

\bibitem{graff2024purposeful}
A.~M. Graff and T.~E. Humphreys, ``Purposeful co-design of {OFDM} signals for
  ranging and communications,'' \emph{{EURASIP} Journal on Advances in Signal
  Processing}, 2024.

\bibitem{bellili2010cramer}
F.~Bellili, N.~Atitallah, S.~Affes, and A.~St{\'e}phenne, ``{Cram{\'e}r-Rao}
  lower bounds for frequency andphase {NDA} estimation from arbitrary square
  {QAM}-modulated signals,'' \emph{IEEE Transactions on Signal Processing},
  vol.~58, no.~9, pp. 4517--4525, 2010.

\bibitem{masmoudi2011non}
A.~Masmoudi, F.~Bellili, S.~Affes, and A.~Stephenne, ``A non-data-aided maximum
  likelihood time delay estimator using importance sampling,'' \emph{IEEE
  Transactions on Signal Processing}, vol.~59, no.~10, pp. 4505--4515, 2011.

\bibitem{zeira1994realizable}
A.~Zeira and P.~M. Schultheiss, ``Realizable lower bounds for time delay
  estimation. 2. {Threshold} phenomena,'' \emph{IEEE transactions on signal
  processing}, vol.~42, no.~5, pp. 1001--1007, 1994.

\bibitem{nanzer2016bandpass}
J.~A. {Nanzer}, M.~D. {Sharp}, and D.~{Richard Brown}, ``Bandpass signal design
  for passive time delay estimation,'' in \emph{2016 50th Asilomar Conference
  on Signals, Systems and Computers}, Nov. 2016, pp. 1086--1091.

\bibitem{sahinoglu2008ultra}
Z.~Sahinoglu, S.~Gezici, and I.~G{\"u}venc, \emph{Ultra-wideband positioning
  systems: theoretical limits, ranging algorithms, and protocols}.\hskip 1em
  plus 0.5em minus 0.4em\relax Cambridge university press, 2008.

\bibitem{barankin1949}
\BIBentryALTinterwordspacing
E.~W. Barankin, ``Locally best unbiased estimates,'' \emph{The Annals of
  Mathematical Statistics}, vol.~20, no.~4, pp. 477--501, 1949. [Online].
  Available: \url{http://www.jstor.org/stable/2236306}
\BIBentrySTDinterwordspacing

\bibitem{mcaulay1971barankin}
R.~McAulay and E.~Hofstetter, ``Barankin bounds on parameter estimation,''
  \emph{IEEE Transactions on Information Theory}, vol.~17, no.~6, pp. 669--676,
  1971.

\bibitem{Ziv1969}
J.~{Ziv} and M.~{Zakai}, ``Some lower bounds on signal parameter estimation,''
  \emph{IEEE Transactions on Information Theory}, vol.~15, no.~3, pp. 386--391,
  1969.

\bibitem{humphreys2023starlinkSignalStructure}
T.~E. Humphreys, P.~A. Iannucci, Z.~M. Komodromos, and A.~M. Graff, ``Signal
  structure of the {Starlink} {Ku}-band downlink,'' \emph{IEEE Transactions on
  Aerospace and Electronic Systems}, pp. 1--16, 2023.

\bibitem{koivisto2017joint}
M.~Koivisto, M.~Costa, J.~Werner, K.~Heiska, J.~Talvitie, K.~Lepp{\"a}nen,
  V.~Koivunen, and M.~Valkama, ``Joint device positioning and clock
  synchronization in {5G} ultra-dense networks,'' \emph{IEEE Transactions on
  Wireless Communications}, vol.~16, no.~5, pp. 2866--2881, 2017.

\bibitem{kakkavas2021power}
A.~Kakkavas, H.~Wymeersch, G.~Seco-Granados, M.~H.~C. Garc{\'\i}a, R.~A.
  Stirling-Gallacher, and J.~A. Nossek, ``Power allocation and parameter
  estimation for multipath-based {5G} positioning,'' \emph{IEEE Transactions on
  Wireless Communications}, vol.~20, no.~11, pp. 7302--7316, 2021.

\bibitem{shamaei2018lte}
K.~Shamaei and Z.~M. Kassas, ``{LTE} receiver design and multipath analysis for
  navigation in urban environments,'' \emph{Navigation}, vol.~65, no.~4, pp.
  655--675, 2018.

\bibitem{shamaei2021receiver}
------, ``Receiver design and time of arrival estimation for opportunistic
  localization with {5G} signals,'' \emph{IEEE Transactions on Wireless
  Communications}, vol.~20, no.~7, pp. 4716--4731, 2021.

\bibitem{neinavaie2021cognitive}
M.~Neinavaie, J.~Khalife, and Z.~M. Kassas, ``Cognitive opportunistic
  navigation in private networks with {5G} signals and beyond,'' \emph{IEEE
  Journal of Selected Topics in Signal Processing}, vol.~16, no.~1, pp.
  129--143, 2021.

\bibitem{neinavaie2023cognitive}
M.~Neinavaie and Z.~M. Kassas, ``Cognitive sensing and navigation with unknown
  {OFDM} signals with application to terrestrial {5G} and {Starlink} {LEO}
  satellites,'' \emph{IEEE Journal on Selected Areas in Communications}, 2023.

\bibitem{xv2023joint}
H.~Xv, Y.~Sun, Y.~Zhao, M.~Peng, and S.~Zhang, ``Joint beam scheduling and
  beamforming design for cooperative positioning in multi-beam {LEO} satellite
  networks,'' \emph{IEEE Transactions on Vehicular Technology}, 2023.

\bibitem{Shi2005}
K.~Shi, E.~Serpedin, and P.~Ciblat, ``Decision-directed fine synchronization in
  {OFDM} systems,'' \emph{IEEE Transactions on Communications}, vol.~53, no.~3,
  pp. 408--412, 2005.

\bibitem{ran2003decision}
J.~Ran, R.~Grunheid, H.~Rohling, E.~Bolinth, and R.~Kern, ``Decision-directed
  channel estimation method for {OFDM} systems with high velocities,'' in
  \emph{The 57th IEEE Semiannual Vehicular Technology Conference, 2003. VTC
  2003-Spring.}, vol.~4.\hskip 1em plus 0.5em minus 0.4em\relax IEEE, 2003, pp.
  2358--2361.

\bibitem{karami2006decision}
E.~Karami and M.~Shiva, ``Decision-directed recursive least squares {MIMO}
  channels tracking,'' \emph{EURASIP Journal on Wireless Communications and
  Networking}, vol. 2006, pp. 1--10, 2006.

\bibitem{xiong2024data}
Y.~Xiong, L.~Tang, S.~Sun, L.~Liu, S.~Mao, Z.~Zhang, and N.~Wei, ``Data-aided
  channel estimation and combining for cell-free massive {MIMO} with
  low-resolution {ADCs},'' \emph{IEEE Communications Letters}, 2024.

\bibitem{ma2001non}
X.~Ma, C.~Tepedelenlioglu, G.~B. Giannakis, and S.~Barbarossa, ``Non-data-aided
  carrier offset estimators for {OFDM} with null subcarriers: identifiability,
  algorithms, and performance,'' \emph{IEEE Journal on selected areas in
  communications}, vol.~19, no.~12, pp. 2504--2515, 2001.

\bibitem{al2006novel}
A.~Al-Dweik, ``A novel non-data-aided symbol timing recovery technique for
  {OFDM} systems,'' \emph{IEEE Transactions on Communications}, vol.~54, no.~1,
  pp. 37--40, 2006.

\bibitem{socheleau2008non}
F.-X. Socheleau, A.~Aissa-El-Bey, and S.~Houcke, ``Non data-aided {SNR}
  estimation of {OFDM} signals,'' \emph{IEEE communications letters}, vol.~12,
  no.~11, pp. 813--815, 2008.

\bibitem{masmoudi2011closed}
A.~Masmoudi, F.~Bellili, S.~Affes, and A.~St{\'e}phenne, ``Closed-form
  expressions for the exact {Cram{\'e}r--Rao} bounds of timing recovery
  estimators from {BPSK}, {MSK} and square-{QAM} transmissions,'' \emph{IEEE
  transactions on signal processing}, vol.~59, no.~6, pp. 2474--2484, 2011.

\bibitem{masmoudi2017nda}
A.~Masmoudi, F.~Bellili, S.~Affes, and A.~Ghrayeb, ``Maximum likelihood time
  delay estimation from single- and multi-carrier {DSSS} multipath {MIMO}
  transmissions for future {5G} networks,'' \emph{IEEE Transactions on Wireless
  Communications}, vol.~16, no.~8, pp. 4851--4865, 2017.

\bibitem{monfared2020iterative}
S.~Monfared, T.-H. Nguyen, T.~Van~der Vorst, P.~De~Doncker, and F.~Horlin,
  ``Iterative {NDA} positioning using angle-of-arrival measurements for {IoT}
  sensor networks,'' \emph{IEEE Transactions on Vehicular Technology}, vol.~69,
  no.~10, pp. 11\,369--11\,382, 2020.

\bibitem{wang2015semiblind}
W.~Wang, T.~Jost, C.~Gentner, S.~Zhang, and A.~Dammann, ``A semiblind tracking
  algorithm for joint communication and ranging with {OFDM} signals,''
  \emph{IEEE Transactions on Vehicular Technology}, vol.~65, no.~7, pp.
  5237--5250, 2015.

\bibitem{adam2013semi}
R.~Adam and P.~A. Hoeher, ``Semi-blind channel estimation for joint
  communication and positioning,'' in \emph{2013 10th Workshop on Positioning,
  Navigation and Communication (WPNC)}.\hskip 1em plus 0.5em minus 0.4em\relax
  IEEE, 2013, pp. 1--5.

\bibitem{mensing2009dd}
C.~Mensing, S.~Sand, A.~Dammann, and W.~Utschick, ``Data-aided location
  estimation in cellular {OFDM} communications systems,'' in \emph{GLOBECOM
  2009 - 2009 IEEE Global Telecommunications Conference}, 2009, pp. 1--7.

\bibitem{laas2021ziv}
T.~Laas and W.~Xu, ``On the {Ziv-Zakai} bound for time difference of arrival
  estimation in {CP-OFDM} systems,'' in \emph{2021 IEEE Wireless Communications
  and Networking Conference (WCNC)}.\hskip 1em plus 0.5em minus 0.4em\relax
  IEEE, 2021, pp. 1--5.

\bibitem{dammann2016optimizing}
A.~Dammann, T.~Jost, R.~Raulefs, M.~Walter, and S.~Zhang, ``Optimizing
  waveforms for positioning in {5G},'' in \emph{2016 IEEE 17th International
  Workshop on Signal Processing Advances in Wireless Communications
  (SPAWC)}.\hskip 1em plus 0.5em minus 0.4em\relax IEEE, 2016, pp. 1--5.

\bibitem{staudinger2017optimized}
E.~Staudinger, M.~Walter, and A.~Dammann, ``Optimized waveform for energy
  efficient ranging,'' in \emph{2017 14th Workshop on Positioning, Navigation
  and Communications (WPNC)}.\hskip 1em plus 0.5em minus 0.4em\relax IEEE,
  2017, pp. 1--6.

\bibitem{graff2024ziv}
A.~M. Graff and T.~E. Humphreys, ``{Ziv-Zakai}-optimal {OFDM} resource
  alocation for time-of-arrival estimation,'' \emph{{IEEE} Transactions on
  Wireless Communications}, 2024, submitted for review{.}

\bibitem{gusi2018ziv}
A.~Gusi-Amig{\'o}, P.~Closas, A.~Mallat, and L.~Vandendorpe, ``{Ziv-Zakai}
  bound for direct position estimation,'' \emph{Navigation}, vol.~65, no.~3,
  pp. 463--475, 2018.

\bibitem{xu2016maximum}
W.~Xu, M.~Huang, C.~Zhu, and A.~Dammann, ``Maximum likelihood {TOA} and {OTDOA}
  estimation with first arriving path detection for {3GPP} {LTE} system,''
  \emph{Transactions on Emerging Telecommunications Technologies}, vol.~27,
  no.~3, pp. 339--356, 2016.

\bibitem{d1994modified}
A.~N. D'Andrea, U.~Mengali, and R.~Reggiannini, ``The modified {Cramer-Rao}
  bound and its application to synchronization problems,'' \emph{IEEE
  Transactions on Communications}, vol.~42, no. 234, pp. 1391--1399, 1994.

\bibitem{bellini1974bounds}
S.~Bellini and G.~Tartara, ``Bounds on error in signal parameter estimation,''
  \emph{IEEE Transactions on Communications}, vol.~22, no.~3, pp. 340--342,
  1974.

\bibitem{bell1997extended}
K.~L. Bell, Y.~Steinberg, Y.~Ephraim, and H.~L. Van~Trees, ``Extended
  {Ziv-Zakai} lower bound for vector parameter estimation,'' \emph{IEEE
  Transactions on information theory}, vol.~43, no.~2, pp. 624--637, 1997.

\bibitem{mehta2007approximating}
N.~B. Mehta, J.~Wu, A.~F. Molisch, and J.~Zhang, ``Approximating a sum of
  random variables with a lognormal,'' \emph{IEEE Transactions on Wireless
  Communications}, vol.~6, no.~7, pp. 2690--2699, 2007.

\bibitem{abramowitz1968handbook}
M.~Abramowitz and I.~A. Stegun, \emph{Handbook of mathematical functions with
  formulas, graphs, and mathematical tables}.\hskip 1em plus 0.5em minus
  0.4em\relax US Government printing office, 1968, vol.~55.

\bibitem{dardari2009ziv}
D.~Dardari and M.~Z. Win, ``{Ziv-Zakai} bound on time-of-arrival estimation
  with statistical channel knowledge at the receiver,'' in \emph{2009 IEEE
  International Conference on Ultra-Wideband}.\hskip 1em plus 0.5em minus
  0.4em\relax IEEE, 2009, pp. 624--629.

\bibitem{chazan1975improved}
D.~Chazan, M.~Zakai, and J.~Ziv, ``Improved lower bounds on signal parameter
  estimation,'' \emph{IEEE transactions on Information Theory}, vol.~21, no.~1,
  pp. 90--93, 1975.

\bibitem{burkhardt2014quadriga}
F.~Burkhardt, S.~Jaeckel, E.~Eberlein, and R.~Prieto-Cerdeira, ``{QuaDRiGa}: A
  {MIMO} channel model for land mobile satellite,'' in \emph{The 8th European
  Conference on Antennas and Propagation (EuCAP 2014)}.\hskip 1em plus 0.5em
  minus 0.4em\relax IEEE, 2014, pp. 1274--1278.

\bibitem{jaeckel20225g}
S.~Jaeckel, L.~Raschkowski, and L.~Thieley, ``A {5G-NR} satellite extension for
  the {QuaDRiGa} channel model,'' in \emph{2022 Joint European Conference on
  Networks and Communications \& 6G Summit (EuCNC/6G Summit)}.\hskip 1em plus
  0.5em minus 0.4em\relax IEEE, 2022, pp. 142--147.

\bibitem{cassiau2020satellite}
N.~Cassiau, G.~Noh, S.~Jaeckel, L.~Raschkowski, J.-M. Houssin, L.~Combelles,
  M.~Thary, J.~Kim, J.-B. Dore, and M.~Laugeois, ``Satellite and terrestrial
  multi-connectivity for {5G}: Making spectrum sharing possible,'' in
  \emph{2020 IEEE Wireless Communications and Networking Conference Workshops
  (WCNCW)}.\hskip 1em plus 0.5em minus 0.4em\relax IEEE, 2020, pp. 1--6.

\bibitem{marshall1960multivariate}
A.~W. Marshall and I.~Olkin, ``Multivariate chebyshev inequalities,'' \emph{The
  Annals of Mathematical Statistics}, pp. 1001--1014, 1960.

\end{thebibliography}
\end{document}